\documentclass[aps,prb, reprint,amsmath,amssymb,longbibliography]{revtex4-2}

\usepackage{graphicx}
\usepackage{dcolumn}
\usepackage{bm}
\usepackage{multirow}
\usepackage{color}
\usepackage{amsmath}
\usepackage{gensymb}
\usepackage[colorlinks=true]{hyperref}
\usepackage{textcomp}
\usepackage{placeins}
\usepackage{braket}
\usepackage[normalem]{ulem}

\usepackage[dvipsnames]{xcolor}
\usepackage{amsmath}
\usepackage{amssymb}
\usepackage{bm}
\usepackage{epsfig}
\usepackage{graphicx}
\usepackage{color}
\usepackage{hyperref}

\def\be{\begin{equation}}
\def\ee{\end{equation}}
\def\bea{\begin{eqnarray}}
\def\eea{\end{eqnarray}}

\begin{document}
	\title{ Cubic anisotropy of hole Zeeman splitting  in semiconductor nanocrystals}
	\author{M. A. Semina}\email{semina@mail.ioffe.ru} \author{A. A. Golovatenko and A. V. Rodina}
	\affiliation{Ioffe Institute, 194021, St.-Petersburg, Russia. }
	\begin{abstract}
We study theoretically cubic anisotropy of  Zeeman splitting of a  hole localized in semiconductor nanocrystal.  This anisotropy originates from three contributions:
crystallographic cubically-symmetric spin and kinetic energy terms in the bulk  Luttinger Hamiltonian  and the spatial wave function distribution in a cube-shaped nanocrystal. 
From  symmetry considerations, an effective Zeeman Hamiltonian for the hole lowest even state is introduced, containing a spherically symmetric and a cubically symmetric term. The values of these terms are calculated numerically for spherical and cube-shaped nanocrystals as functions of the Luttinger Hamiltonian parameters. We demonstrate that the  cubic shape of the nanocrystal and the cubic anisotropy of hole kinetic energy (so called valence band warping) significantly affect effective $g$ factors of  hole states. In both cases, the effect comes from the cubic symmetry of the hole wave functions in zero magnetic field.  Estimations for the effective $g$ factor values in several semiconductors with zinc-blende crystal lattices are made. Possible experimental manifestations and potential methods of measurement of the cubic anisotropy of the hole Zeeman splitting  are suggested.

	\end{abstract}	
	\maketitle
\date{\today}


\section{Introduction}

Today, state-of-the-art methods of  chemical synthesis  make it possible to grow  semiconductor colloidal quantum dots or nanocrystals (NCs) with desired shapes on demand. For example, typical II-VI semiconductor (CdS, CdSe, CdTe) NCs tend to have a shape close to spherical \cite{Efros2021} and lead-halide perovskite NCs naturally grow in the form of cubes \cite{Kovalenko2015}. However, by choosing special synthesis conditions, it is possible to control the shape of NCs.
It is known that CdSe NCs can be obtained in the form of nanorods,  nanoplatelets,  tetrapods \cite{Efros2021}. Recently, cube-shaped CdSe and CdS NCs with zinc-blende crystal structure have been synthesized \cite{Lv2022}. This diversity of NC shapes is interesting not only from the point of view of the capabilities of modern methods of colloidal synthesis, but also because it opens the possibility of studying the effect of the NC shape on the physical properties of charge carriers localized in NCs.

The dependence of  size quantization energy on NC shape is known from textbooks,  but there are more subtle effects, for example, the strong dependence of the spin properties of localized charge carriers on NC size and symmetry. The enduring interest in the spin properties of charge carriers localized in NCs is associated with the possibility of their application in spintronics and quantum computing, as was proposed in the Loss-DiVincenzo paper \cite{Loss1997}. The main searches in this direction are based on the use of direct band gap semiconductors of type III-V (GaAs) or II-VI (CdSe, CdTe), which have zinc-blende crystal structure.

These semiconductor materials are characterized by a strong spin-orbit interaction. For electrons in the $S$-type conduction band, the spin-orbit interaction results in deviation of the electron $g$ factor from the free electron $g$ factor, as it was shown by Lora Roth for bulk semiconductors \cite{Roth1959}. In small NCs, spatial localization of electrons decreases the spin-orbit interaction effect on electron $g$ factor, so that its value  tends to the free electron $g$ factor  \cite{Kiselev1998,Tadjine2017,Semina2021}. In spherically and cubically-symmetric localizing potentials, the splitting of electron spin sublevels is described by  an isotropic $g$ factor.

The topmost valence band in these materials has a $P$-type symmetry. Due to the strong spin-orbit interaction, the valence band splits into two subbands with the total angular momentum of a hole $J=3/2$ and $J=1/2$. The splitting between these subbands reaches hundreds of meVs. The ground hole state in these materials corresponds to the four-fold degenerate valence subband with $J=3/2$. The calculation of the hole $g$ factor in this case within the $\bm {kp}$-method for bulk semiconductors was first performed by Luttinger \cite{Luttinger1956}. He showed that the cubic symmetry of the crystal lattice leads to two additional contributions to the splitting of hole states in a magnetic field. The first contribution is due to the valence band warping, and the second to the cubic-symmetric invariant $\propto J^3B$. These contributions results in dependence of the hole spin splitting in an  external magnetic field on the angle between crystal axes and magnetic field. 
The influence of these contributions on a hole localized at different types of acceptors in GaAs was studied in Ref. \cite{Malyshev2000}. 
It was found that the contribution to the hole spin splitting from the cubically symmetric term of the Luttinger Hamiltonian can be comparable to the spherically symmetric contribution \cite{Malyshev2000}. 

However, often the contribution originating from the cubic symmetry of the crystal lattice is neglected, and the hole $g$ factor is calculated in the spherical approximation of the Luttinger Hamiltonian \cite{Gelmont1973,Semina2015}. This approach was widely used for  interpretation of magneto-optical experiments on colloidal NCs \cite{efros1996,JohnstonHalperin2001,Htoon2009,Biadala2010,Liu2013,Sinito2014}. To the best of our knowledge, the applicability of such an approach has not been assessed yet. Since the cubically-symmetric contribution to the  effective $g$ factor  of a hole localized on an acceptor \cite{Malyshev2000} is not negligible, one can expect the similar result for a hole localized in a NC.

For instance, accounting for the cubic symmetry of a crystal lattice becomes crucial for understanding  the spin properties of holes in thin quantum wells. 
Due to a large energy splitting of the  localized  states of the light and heavy holes in thin quantum wells, the transverse $g$ factor of the heavy hole turns out to be identically equal to zero if one neglects the cubic symmetry of the crystal lattice. As a result, no spin precession of a heavy hole should be observed, which is not the case \cite{Marie1999,Syperek2007,Kugler2009}. 
It was shown in \cite{Marie1999} that accounting for the crystal lattice cubic symmetry  in the form of the Hamiltonian term $\propto J^3B$ results in the heavy hole having a nonzero transverse $g$ factor, which explains experimental findings. 

The effect of cubic symmetry on the spin splitting of a localized hole originates not only from the crystal lattice, but also from the localization potential symmetry \cite{Semina2021}. Such a localization potential does not exist naturally in bulk semiconductors, but has been realized recently in cube-shaped CdSe NCs \cite{Lv2022}. Thus, it is interesting to reveal the effect of NC shape on hole spin splitting by comparing it in spherical and cube-shaped NCs. 

In this paper, we study theoretically the spin splitting of holes in NCs of spherical and cubic shape in an applied magnetic field. We analyze and compare contributions of both the cubic symmetry of a semiconductor crystal structure and a NC's shape to hole spin splitting in an external magnetic field. It is found that in cube-shaped NCs,  both contributions to the effective hole $g$ factor are comparable with each other, and, although smaller, not negligible as compared to the isotropic contribution. We propose possible experimental manifestations of the cubically-symmetric contribution to the hole spin splitting. 

The paper is organized as follows. In section II we introduce the form of the hole effective Zeeman Hamiltonian in an external magnetic field from symmetry consideration. In sections III and IV we present results of calculations of hole Zeeman splitting without accounting for valence band warping in spherical and cube-shaped NCs. In section V, we analyse the contribution from  valence band warping on NC of both shapes. In section VI, we discuss possible experimental manifestations of the cubic anisotropy of hole Zeeman splitting.

\section{Symmetry consideration}

We consider a hole from the top of the valence band in NCs based on semiconductors with a zinc-blende crystal lattice and a strong spin-orbit interaction. 
For a hole localized in an external potential $V_{\text{ext}}(\bm r)$, and in the presence of an applied magnetic field, the Hamiltonian takes form:
\begin{equation}\label{Hamilt}
\widehat{H}=\widehat{H}_L+\widehat{H}_{\text{Z}}+\widehat{H}_{\text{B}}+V_{\text{ext}}(\bm r).
\end{equation}

The dispersion of hole states in the top of the valence band is described  by  the four-band Luttinger Hamiltonian $\widehat{H}_L$ \cite{Luttinger1956,Gelmont1971}:
\begin{widetext}\begin{equation}\label{lutt}
\widehat{H}_L=\frac{\hbar^2}{2 m_0}\left[\left(\gamma_1+\frac{5}{2}\gamma_2\right)k^2-2\gamma_2\sum_{\alpha}J_{\alpha}^2k_{\alpha}^2-2\gamma_3\sum_{\alpha\neq\beta}\{J_\alpha J_\beta\}\{k_{\alpha}k_{\beta}\}\right],\quad \alpha,\beta=x,y,z.
\end{equation}
\end{widetext}
Here, the operator ${\bm J}$ is the hole internal angular momentum operator,  for $\Gamma_8$ the valence subband $J=3/2$,  $\gamma_1,\gamma_2,\gamma_3$  are Luttinger parameters, $k_{\alpha}$ are the hole wave vector components, and  $\{ab\}=(ab+ba)/2$. The components of the wave vector $k_{\alpha}$  and pseudovector $J_\alpha$ transform according to the irreducible representation $\Gamma_8$ of the point group $O_h$ or $T_d$.  The Luttinger Hamiltonian \eqref{lutt} comprising three cubically-symmetric invariants is written for semiconductors with cubic symmetry of the crystal lattice with an inversion center. The effects of  odd linear and cubic in $k_\alpha$  terms, which are  allowed if the inversion center is lacking, for example, in the $T_d$ point group corresponding to a zinc-blende crystal lattice, are not discussed in this work. 

The external magnetic field gives two contributions to the hole Hamiltonian \eqref{Hamilt}. The first one is the Zeeman contribution, $\widehat{H}_{\text{Z}}$ \cite{Luttinger1956, Winkler2003}: 
\begin{multline}\label{HZ}
\widehat{H}_{\text{Z}}=-2\mu_B\varkappa\left(\bm J\bm B\right)-\\-2\mu_B q\left(J_x^3B_x+J_y^3B_y+J_z^3B_z\right),
\end{multline}
with $\varkappa$ and $q$ being magnetic Luttinger parameters \cite{Luttinger1956}. 
The first term in \eqref{HZ} $\propto \varkappa$ is spherically symmetric and the  value of the parameter $\varkappa$ can be estimated using perturbation theory as \cite{Roth1959}
\begin{equation}\label{kappa}
\varkappa\approx-2/3+(2\gamma_2+3\gamma_3)/3-\gamma_1/3.
\end{equation} 
The second term in \eqref{HZ}  $\propto q$  has cubic symmetry, with axes  naturally coinciding with the crystal lattice axes. It is often neglected as the value of $q$ is small in typical semiconductors being  two orders of magnitude less than $\varkappa$ \cite{Winkler2003,Marie1999}. The matrix form of $\hat{H}_Z$ \eqref{HZ}, calculated in  the standard basis of the topmost $\Gamma_8$ valence subband\cite{Bir1974book,Ivchenko2005book} is shown in Appendix \ref{AA}, Eq. \eqref{Hzmatrix}. 

The second contribution from the magnetic field is the orbital contribution, $\widehat{H}_{\text{B}}$. It comes from the hole wave vector  ${\bm k}$  in the Luttinger Hamiltonian  $\widehat{H}_L$ \eqref{lutt} being replaced  by  ${\bm k} -\frac{e}{c}\bm A$, where  $\bm A$ is the vector potential of the electro-magnetic field.
We consider an arbitrary directed magnetic field and use the symmetric gauge 
\begin{multline}\label{guage}
\bm A=\frac{1}{2}[\bm B\times \bm r]=\\=\frac{1}{2}(B_yz-B_zy,B_zx-B_xz, B_xy-B_yx).
\end{multline} 
The given selection of $\bm A$ is necessary for studying of the hole $g$ factor anisotropy. This task can not be solved if one considers  $\bm B|| z$ and $\bm A = (0,Bx, 0)$ \cite{Semina2016,Semina2015,Semina2021}.

The explicit form of the orbital contribution  $\widehat{H}_{\text{B}}$ for the gauge  \eqref{guage}, neglecting the quadratic on magnetic field terms is shown in Appendix \ref{AA}, Eqs. \eqref{HBmatrix_sph} and \eqref{HBmatrix_warp}.
Note, that the Luttinger Hamiltonian $\widehat{H}_L$  has   cubic symmetry  due to valence band warping, $\gamma_2\neq\gamma_3$. As a result, the  magnetic field orbital contribution   $\widehat{H}_{\text{B}}$ also inherits cubic symmetry in the case $\gamma_2\neq\gamma_3$. Both  $\widehat{H}_L$  and $\widehat{H}_{\text{B}}$  can be  separated into isotropic and cubically symmetric parts, see Eqs. \eqref{lutt_matrix_sph}, \eqref{lutt_matrix_warp}, \eqref{HBmatrix_sph}, and \eqref{HBmatrix_warp}.  

In a weak magnetic field, the effect of the magnetic field on hole states can be calculated using first order perturbation theory by considering $\widehat{H}_{\text{Z}}$ and $\widehat{H}_{\text{B}}$ as perturbations.  In the framework of the four-band Luttinger Hamiltonian \eqref{lutt}  neglecting the admixing of hole states from spin-orbit split-off valence subband,  hole effective $g$ factors would not depend on an NC size, but only on its shape and the type of localizing potential \cite{Semina2015, Semina2021}.  As the value of the hole effective $g$ factor is very sensitive to the wave function structure, one first has to find  proper zero field wave functions for $\widehat{H}_{\text{L}}+V_{\text{ext}}(r)$. For our calculations, we  use numerical methods developed in \cite{Semina2015, Semina2021} and evolved here.

In zero magnetic field, and in bulk semiconductors for $V_{\text{ext}}(\bm r)\equiv 0$, the hole states at the top of the valence band (${\bm k}=0$) are four-fold degenerate. 
The perturbation $\hat H_{\text Z}$ lifts this degeneracy completely  in a weak external magnetic field. Without accounting for the cubically-symmetric contribution ($q=0$), this splitting is isotropic and equidistant. The four spin states are characterized by the projection of ${\bm J}$ on the magnetic field, and the splitting between neighboring states is equal to $2 \varkappa \mu_B B$. The non-zero cubically-symmetric contribution $\propto q \ne 0$ makes the hole Zeeman splitting anisotropic (depending on the angle between the magnetic field direction and crystallographic axes) and non-equidistant. If the magnetic field is directed along one of the crystal axes denoted as $z$,  the splitting becomes   $E_{-3/2}-E_{+3/2}=3 g_{h,3/2} \mu_B B$ and $E_{-1/2}-E_{+1/2}=g_{h,1/2} \mu_B B$, where     $g$ factors of heavy ($J_z=\pm 3/2$) and light ($J_z=\pm 1/2$)  holes are:
\begin{equation}\label{g_bulk}
g_{h,3/2}=2\varkappa+\frac{9}{2}q,\quad g_{h,1/2}=2\varkappa+\frac{1}{2}q,
\end{equation}
Note, that values of $g_{h,3/2}$ and $g_{h,1/2}$ are different, although this difference is small due to a small value of $q$ \cite{Marie1999}.  In bulk semiconductors, the contribution $\widehat H_B$ becomes important at high magnetic fields, when  hole Landau levels are formed \cite{Luttinger1956}. In weak magnetic fields, $\widehat{H}_{\text{B}}$ does not affect the hole $g$ factor.

The situation changes for localized holes, both in zero and external magnetic fields. The symmetry of the localized hole states depends additionally on the symmetry of the external localizing potential $V_{\text{ext}}(\bm r)$, which can, in general, be arbitrary. In this work, we consider $V_{\text{ext}}(\bm r)$ with the spherical or cubic symmetry. In the case of a cubically symmetric  $V_{\text{ext}}$, it is assumed that the cubic axes of the localizing potential coincide with the crystal lattice axes. As a result, the symmetry of the whole system is spherical or  cubic. 

In spherically-symmetric systems, when cubically-symmetric  corrections can be neglected, the hole states are classified by their total angular momentum ${\bm j}={\bm J} +{\bm l}$ (${\bm l}$ is the orbital angular momentum) and its projection $j_z=M$ on some marked out axis \cite{Gelmont1971,Baldereschi1973}. Due to  inversion symmetry, the states with even and odd values of $l$ are separated and the ground states is in most cases even. In cubic crystals, hole states have to be classified by irreducible representations of the corresponding point group, in our case $O_h$. Therefore, both the internal angular momentum and the total angular momentum of the hole, as well as their projections, are no longer good quantum numbers \cite{Baldereschi1974}. However, the hole ground state in zero magnetic field remains four-fold degenerate and can be traced to the spherically symmetric state with total angular momentum $j=3/2$.  It can be described by  cubic invariants constructed from the pseudovector components $j_\alpha=J_\alpha + l_\alpha$ ($\alpha=x,z,y$) transforming like   $J_\alpha$  according to the  $\Gamma_8$ representation. Pseudospin $\bm j$ can be considered as a generalized hole angular momentum. As we neglect the odd in wave vector contributions to the hole Hamiltonian, all  results presented below are valid both for $O_h$ and $T_d$ point groups, so for brevity, we omit the parity indices. We further consider  even hole states with $j=3/2$ and projections $M=j_z=\pm 3/2, \pm 1/2$ on the crystallographic axis $z$. Importantly, for this state, the matrices $j_\alpha$ in the basis of the eigenstates $\Psi_M$ of the zero-field Hamiltonian $\hat H_L+V_{\rm ext}$ have the same form as the matrices $J_\alpha$ in the basis of four Bloch functions $u_\mu$, $\mu=\pm 3/2, \pm 1/2$ of  $\Gamma_8$ valence band with  $J_z=\mu$ \cite{ivchenko05a}.  

For $V_{\text{ext}}(\bm r)\neq 0$ the orbital contribution from the magnetic field, $\widehat{H}_{\text{B}}$, becomes non-zero.  In contrast to bulk, for a localized hole two  more factors may result in the cubically-symmetric corrections to the Zeeman splitting: the cubic shape of the NC coming from $V_{\text{ext}}(\bm r)\neq 0$ itself, and  valence band warping.  Both these factors lead to a hole wave function of cubic symmetry in zero magnetic field. We study the effect of these factors in addition to the $\propto q$ cubic term in the Zeeman splitting of the hole states in an external magnetic field.


In an external magnetic field, the effective Zeeman Hamiltonian of a localized hole with $j=3/2$ can be written as \cite{Malyshev2000}:
\begin{multline}\label{effect_Z}
\hat{H}_Z^{\text{eff}}= -\bm \mu^{\rm eff} {\bm B} =  \\ -\mu_B g_{\rm h}\left(\bm j\bm B \right)-2\mu_B Q^{\text{eff}}\left(j_x^3B_x+j_y^3B_y+j_z^3B_z\right),
\end{multline}
where $g_{\rm h}$ and $Q^{\text{eff}}$ are  responsible for the isotropic and cubically-symmetric parts of the hole effective magnetic moment ${\bm \mu}^{\rm eff}$, correspondingly. In the general case, ${\bm \mu}^{\rm eff}$ is not co-linear to ${\bm j}$ and their relation is :
\begin{equation}\label{mueff}
\mu_{\alpha}^{\rm eff}=\mu_B g_{\rm h}j_{\alpha}+2\mu_BQ^{\text{eff}}j_{\alpha}^3.
\end{equation}
The first term linear in $j$ in \eqref{mueff} should be written with the help of a second rank tensor as $g^h_{\beta \alpha}j_\beta$, and the second one, cubic in $j$,  with the help of a fourth rank tensor $Q_{\beta \gamma \delta  \alpha}j_\beta j_\gamma j_\delta$.   In $O_h$ and $T_d$ symmetry, a second rank tensor is diagonal and described by one scalar $g_{\rm h}^{sph}=g^h_{xx}=g^h_{yy}=g^h_{zz}$, while the fourth rank tensor $Q_{\beta \gamma \delta  \alpha}$ is described by two non-zero constants $Q_1=Q^{\rm eff}_{\alpha \alpha \alpha \alpha}$, $\alpha=x,y,z$ and $Q_2=Q^{\rm eff}_{\alpha \alpha \beta \beta}=Q^{\rm eff}_{\alpha \beta \beta \alpha}=Q^{\rm eff}_{\alpha \beta \alpha \beta}$ with $\alpha\neq\beta$, $\alpha,\beta=x,y,z$.  However, as the $j_x^2+j_y^2+j_z^2$ transforms as a scalar in $O_h$ and $T_d$, the two cubically-symmetric contributions can be reduced to one isotropic contribution $\propto Q_2$ added to the $g_h^{\rm sph}$ term and one cubically-symmetric contribution, so that
\begin{equation}\label{Qeff}
g_{\rm h}=g_h^{\rm sph}+41Q_2/2, \quad Q^{\rm eff} = Q_1-3Q_2 \, .
\end{equation}
Deriving Eqs.  \eqref{mueff} and \eqref{Qeff} we took into account $j_x^2+j_y^2+j_z^2 = j(j+1)=15/4$ for $j=3/2$ and computational relations for $j_\beta j_\gamma - j_\gamma j_\beta =i e_{ \beta \gamma \delta} j_\delta$, where $e_{\gamma \beta \delta}$ is a third rank anti-symmetric unit tensor.  

\begin{figure*}
\includegraphics[width=1\textwidth]{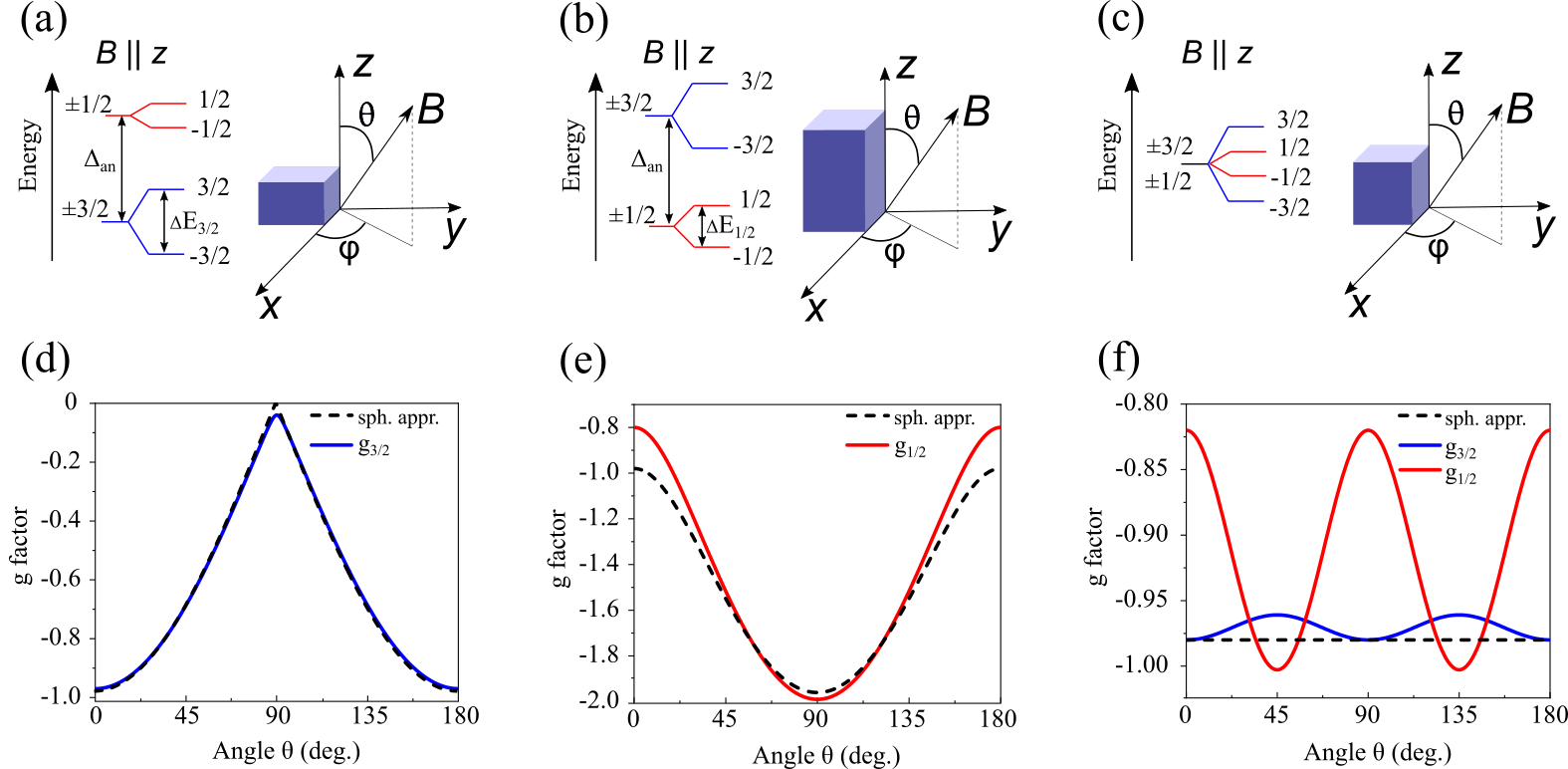}
   \caption{ Schematic of the relative orientation of a NC, applied magnetic field $B$, and the laboratory frame $xyz$ for an oblate NC (a), prolate NC (b), and cube-shaped NC (c). The schematic shows the splitting of hole states in the presence of uniaxaial anisotropy $\Delta_{\rm an}$ and applied magnetic field for the shown NC shapes;  (d-f)  dependencies of the effective $g$ factors $g_{3/2}=\Delta E_{3/2}/3\mu_BB$  and $g_{1/2}=\Delta E_{1/2}/\mu_BB$ of the  hole ground state on the angle $\theta$ between magnetic field direction and $z$. Calculations are done for $\varphi=0$, $g_h=-0.8$, and $Q^{\text{eff}}=-0.04$. Dashed line corresponds to $g_h=g_h^{\rm sph}=-0.98$ and $Q^{\text{eff}}=0$. 
   } 
 \label{g_angle}
\end{figure*}

If the symmetry of $V_{\rm ext}$ is additionally lowered to include some uniaxial  perturbation $V_{\rm ext}^{\rm an}$ along crystal axis $z$ with corresponding point symmetry group $D_{4h}$, the hole states with $|M|=3/2$ (heavy holes) and $|M|=1/2$ (light holes) are  split in zero magnetic field, as  shown in Fig. \ref{g_angle}(a,b).  Then, the axes $z$ and $x,y$ become nonequivalent, and the hole effective Hamiltonian in an external magnetic field should be written as  
\begin{multline}\label{effect_Z_D4h}
\hat{H}_Z^{\text{eff}}=  -\mu_B \left[g_{\perp}(j_xB_x+j_yB_y)+g_{\parallel}j_zB_z \right]-\\
-2\mu_B \left[Q^{\text{eff}}_{\perp}(j_x^3B_x+j_y^3B_y)+Q^{\text{eff}}_{\parallel}j_z^3B_z\right].
\end{multline}
Instead of four parameters $g_{\perp}$, $g_{\parallel}$, $Q^{\text{eff}}_{\perp}$ and $Q^{\text{eff}}_{\parallel}$, it is instructive to introduce the following four parameters
\begin{multline}\label{gfactors_cube_2}
g_{3/2}^{||}=g_{\parallel}+\frac{9}{2}Q^{\text{eff}}_{\parallel},\quad g_{1/2}^{||}=g_{\parallel}+\frac{1}{2}Q^{\text{eff}}_{\parallel},\\
g_{3/2}^{\perp}=Q^{\text{eff}}_{\perp},\quad g_{1/2}^{\perp}=2g_{\perp}+10Q^{\text{eff}}_{\perp}.
\end{multline}
If the splitting energy $\Delta_{an}$ between the  heavy  ($|M|=3/2$)  and light ($|M|=1/2$) hole states is large  as compared with the magnetic field induced splitting, $\Delta_{an} \gg \mu_B B$, in the magnetic field they can be considered separately. 
In this case, parameters of \eqref{gfactors_cube_2} have the meaning of the effective longitudinal (transverse) $g$ factors describing the splitting of heavy and light hole states in a magnetic field directed parallel (perpendicular) to the anisotropy axis. 
For an arbitrary field direction, the Zeeman splitting of heavy and light  holes, in this case, would be described by the matrices Eq. \eqref{HZ32} and Eq. \eqref{HZ32} given 
in Appendix \ref{AA}.  
It is isotropic in $xy$ plane and highly anisotropic with respect to the angle $\theta$ between the magnetic field and $z$ axis:
\begin{equation}\label{Zeeman_simple}
\Delta E_{|M|}=2|M|\mu_B\sqrt{(g_{|M|}^{||}B_z)^2+(g_{|M|}^{\perp}B_{\perp})^2}.
\end{equation} 
where $B_{\perp}=\sqrt{B_x^2+B_y^2} = B \sin \theta$.  

Importantly, 
 for $B_z=0$, the Zeeman splitting of holes with $|M|=3/2$ is not zero only due to a non-zero $g_{3/2}^{\perp}=Q^{\text{eff}}_{\perp}$, while the difference between $g_{3/2}^{||}$ and $g_{1/2}^{||}$ arises only due to a nonzero $Q^{\text{eff}}_{||}$. If the cubically-symmetric terms are absent in the Luttinger Hamiltonian and external potential, $Q^{\text{eff}}_{\perp} = 0$, the transverse heavy hole $g$ factor vanishes. At the same time, the uniaxial symmetry of $V_{\rm ext}^{\rm an}$ , for example, for spheroidal NCs,  contributes to $g_{\perp} 
 \ne g_{\parallel} \ne g_{\rm h}^{\rm sph}$ and induce $Q^{\text{eff}}_{\parallel} \ne 0$. This results in $g_{3/2}^{||} \ne g_{1/2}^{||}$ in spheroidal NCs, as shown in \cite{Semina2021}. 
 
As we study NCs with shape close to spherical or cubic in what follows, we focus on  effects stemming from  a non-zero cubically-symmetric contribution to the effective Zeeman Hamiltonian \eqref{effect_Z}. We neglect the effect of $V_{\rm ext}^{\rm an}$ on the uniaxial anisotropy of hole Zeeman splitting and assume  in Eqs. \eqref{effect_Z_D4h}  and \eqref{gfactors_cube_2} that
\begin{equation}
g_{\perp} \equiv g_{\parallel}\equiv g_{\rm h},\quad   Q^{\text{eff}}_{\perp} \equiv Q^{\text{eff}}_{\parallel}\equiv Q^{\text{eff}}.
\end{equation}
The effective $g$ factors dependencies on the angle $\theta$, for hole states  with $|M|=3/2$ and $|M|=1/2$, calculated accounting for the cubic symmetry of the crystal lattice for parameters of CdSe, $g_h=-0.8$ and $Q^{\text{eff}}=-0.04$ are shown  in Figs.~\ref{g_angle}(d,e), respectively. Dashed lines in Fig.~\ref{g_angle}(d,e) show the effective $g$ factor calculated neglecting lattice symmetry with $g_h=-0.98$ and $Q^{\text{eff}}=0$.

If the hole states with $|M|=3/2$ and $|M|=1/2$ are degenerate in zero magnetic field, the matrix form of the effective Zemman Hamiltonian is given in Eq. \eqref{Hzeff}, Appendix \ref{AA}. 
In this case, a simple expression like Eq. \eqref{Zeeman_simple} for the hole Zeeman splitting in an arbitrary directed magnetic field cannot be written. For the magnetic field directed along a crystal axis, the four hole states in the magnetic field are characterized by the projection  $M_B=  \pm 3/2, \pm 1/2$ of ${\bm j}$ on ${\bm B}$, $M_B=M$ if ${\bm B}\parallel z$. For an arbitrary field direction, the $\propto Q^{\rm eff}$ term mixes the states with different $M_B=({\bm j} {\bm B} )/B$ as well as with different $M=j_z$. However, we continue to notate the four eigenstates and their respective Zeeman splitting obtained by the  diagonalizaion of the matrix Eq. (A3) with $M_B=\pm 3/2, \pm 1/2$ and 
$$\Delta E_{3/2} = \frac{3}{2} g_{3/2} \mu_B B, \, \, \Delta E_{1/2} = \frac{1}{2} g_{1/2} \mu_B B.$$
 In that case, the hole Zeeman splittings are anisotropic with all cubic axes being equal.  In Fig. \ref{g_angle}(f), we show dependence of the effective $g$ factor on the  angle $\theta$ with the same set of parameters as in Fig.~\ref{g_angle}(d,e). 

As stated above, the hole Hamiltonian \eqref{Hamilt} contains three sources of the cubically-symmetric contribution to the hole $g$ factor: a crystallographic contribution to the Zeeman Hamiltonian $\propto q$, the shape of the NC coming in form of $V_{\text{ext}}(\bm r)$, and valence band warping,  $\gamma_2\neq\gamma_3$. Below we show how these three factors contribute to $g_{\rm h}$ and $Q^{\text{eff}}$. It is important to note that the values of $g_h$ and $Q^{\rm eff}$ can be determined from calculations performed for the magnetic field along one of the cubic axes, as they are equal. However, we verify the resulting $g$ factors and Zeeman splittings by direct numeric calculations for an arbitrary direction of magnetic field.

We start from the first two effects and consider in Sections \ref{sec_sph} and \ref{sec_cub} the  Hamiltonian \eqref{lutt} in the spherical approximation neglecting  valence band warping:
\begin{equation}\label{lutt_sph}
\widehat{H}_L=\frac{\hbar^2}{2 m_0}\left[\left(\gamma_1+\frac{5}{2}\gamma\right)k^2-\gamma\{J_\alpha J_\beta\}\{k_\alpha k_\beta\}\right],
\end{equation}
where $\gamma=(2\gamma_2+3\gamma_3)/5$ and the bulk light-hole and heavy-hole effective masses are: $m_{lh} = m_0/(\gamma_1 + 2 \gamma)$ and $m_{hh} = m_0/(\gamma_1 - 2\gamma)$, respectively. In Sec. \ref{sec_wap} we consider additionally the effect of the $\gamma_2\neq\gamma_3$. We demonstrate that the Zeeman crystallographic cubically-symmetric contribution $\propto q$ contributes only to the $Q_1$ constant and, thus, to $Q^{\text{eff}}$, but does not change the isotropic part of hole effective $g$ factor.  The cubic shape of NCs as well as the cubic terms in the hole kinetic energy, on the contrary, contribute to both $Q_1$ and $Q_2$, hereby changing isotropic part  of hole $g$ factor, $g_h$, as well as inducing the $Q^{\text{eff}}$ term even for $q=0$.  

\section{Results and discussion}

\subsection{Spherically-symmetric nanocrystals} \label{sec_sph}

For spherically symmetric NCs considered in the spherical approximation of the Luttinger Hamiltonian \eqref{lutt_sph}, the  cubically-symmetric contribution to the hole $g$ factor may originate only  from the term  $\propto q$ in $\hat H_Z$.   
In a spherically symmetric external potential ${V}_{\text{ext}}({\bf r})$ with isotropic  kinetic energy, hole  states can be classified by their total angular momentum ${\bm j}$ \cite{Gelmont1971, Baldereschi1973,SercelPRL90, SercelPRB90} and their wave functions are   \cite{Gelmont1971}:
\begin{multline}
\Psi_{jM} = \sqrt{2j+1}\sum_{l} (-1)^{l-3/2+M} (i)^l R_{jl}(r)\times \\ \times\sum_{m+\mu = M}
\left(
\begin{array}{ccc}
l & 3/2&j \\ m&\mu&-M
\end{array}
\right)
Y_{l,m} u_\mu \, ,
\label{Gelmont_functions}
\end{multline}
Here, $Y_{lm}$ are spherical harmonics \cite{Edmonds} being the eigenfunctions of the hole orbital momentum $l$, $\left(_{m~n~p}^{i~~k~~l}\right)$ are $3j$ Wigner symbols.
The ground hole state is a four-fold degenerate $SD$-like  state and consists of functions with $l=0$ and $l=2$. For simplicity, we denote the respective radial functions as $R_0(r)$ and $R_2(r)$.

\begin{figure}
\includegraphics[width=0.99\columnwidth]{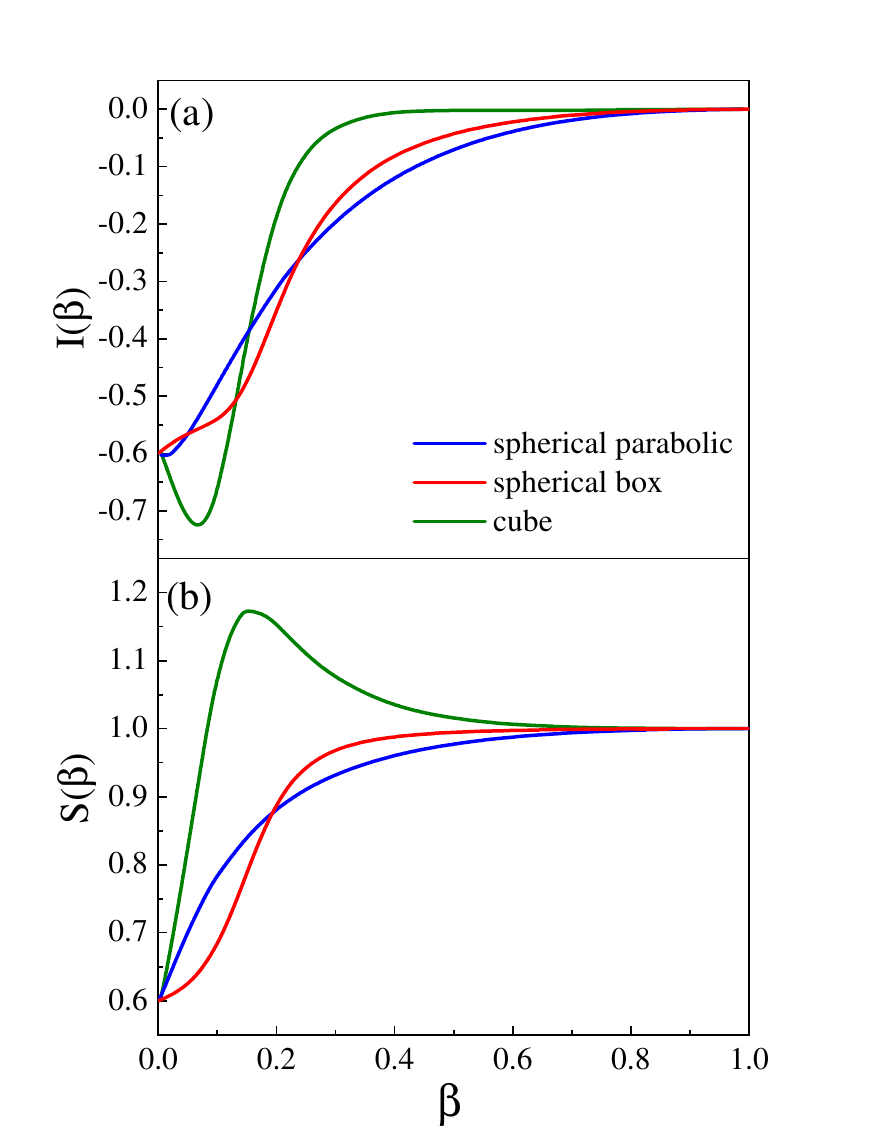}
   \caption{Dependencies of the functions (a) $I(\beta)$ and (b) $S(\beta)$ on the light to heavy hole effective mass ratio $\beta$ for spherical NCs with parabolic and box-like potentials and cube-shaped NCs.   }
 \label{fig:1}
\end{figure}

For $q\equiv 0$, the hole $g$ factor  was calculated in  Ref. [\onlinecite{Gelmont1973}] and can be written as (see Eq. (12) of Ref. [\onlinecite{Semina2021}] in the limit of strong spin-orbit interaction) 
\begin{eqnarray}\label{Gelmontgen}
&&	g_{\rm h}^{\text{sph}}  = 2\varkappa S(\beta) + \frac{4}{5} \gamma_1 I(\beta)  \, ,  \\
&&	S(\beta) = \left(1- \frac{4}{5} I_2^{\text{g}} \right) \, , \quad 
	I(\beta) = \frac{1-\beta}{1+\beta} I_1^{\text{g}}+ \frac{2\beta}{1+\beta} I_2^{\text{g}} \, ,\nonumber
	\end{eqnarray}
where $\beta$ is the light to heavy hole effective mass ratio $\beta=(\gamma_1-2\gamma)/(\gamma_1+2\gamma)$. Function $S(\beta)$ describes the renormalization of the isotropic part of the Zeeman contribution $\widehat H_{Z}$,  function $I(\beta)$ describes the orbital contribution to the $g$ factor stemming from $\widehat H_B$.   
Integrals $I_1^{\text{g}}$ and $I_2^{\text{g}}$ were introduced in Ref. [\onlinecite{Gelmont1973}]:
\begin{equation}\label{Gelmont1}
I_1^{\text{g}}=\int\limits_{0}^{\infty}r^3 R_2(r) \frac{d R_0(r)}{ dr} dr,~ I_2^{\text{g}}=\int\limits_{0}^{\infty}r^2 R_2^2(r) dr .
\end{equation}
The dependencies  $I(\beta)$ and $S(\beta)$ for a spherically-symmetric NC with parabolic and box-like potentials are shown in Fig. \ref{fig:1}.

If $q\neq 0$, the corresponding contribution to the hole $g$ factor can be calculated using first order perturbation theory. 
It can be easily shown, that for a hole  wave function in the form of Eq. \eqref{Gelmont_functions}, $Q_2=0$, $g_h=g_h^{\rm sph}$ and $Q^{\rm eff}=Q_1=Q^q$, where  function $Q^q(\beta)$ can be written as
\begin{equation}\label{Qq}
Q^q(\beta)=\left(1- \frac{4}{5} I_2^{\text{g}} \right)=S(\beta). 
\end{equation}
One can see in Figure \ref{fig:1} that $Q^q(\beta)=S(\beta)$  does not exceed  unity in the whole range of $\beta$. Thus, in spherical NCs, the contribution to the hole $g$ factor from the cubically symmetric part of the Zeeman Hamiltonian is comparable to the corresponding contribution in bulk.

\subsection{Cube-shaped nanocrystalls}\label{sec_cub}

In cube-shaped NCs there is an additional source of the cubic symmetry arising from the shape of the NC. As an example, we consider cube-shaped NCs  with rectangular infinite localization potential.   It was shown that  the hole ground state is an even $S$-like state for the majority values of  $\beta$, although, for $0.1425\gtrsim\beta\gtrsim 0.0775$ the ground state for cube-shaped NCs with  box-like infinite potential is a $P$-like odd state \cite{Semina2021}. Both states are four-fold degenerate, as the cubically symmetric terms in the Luttinger Hamiltonian do not split states with total angular momentum less than $5/2$ \cite{Baldereschi1974}.  Here we consider only the lowest even hole state, which can be traced to the $S_{3/2}$ state in spherical NCs. 

As we have shown in Ref. [\onlinecite{Semina2021}], even if we neglect the cubic contribution to the Zeeman Hamiltonian \eqref{HZ},  and consider the  Luttinger Hamiltonian in the spherical approximation, $g$ factors of ``heavy'' ($|j_z|$=3/2) and ``light'' ($|j_z|=1/2$) holes are different in cube-shaped NCs.    In Ref. [\onlinecite{Semina2021}] we qualitatively explained this result by an effective cubically symmetric contribution to the hole $g$ factor coming  from the NC shape. In this work, we study the effect of the cubic shape of the NC in more detail.

Unlike the wave functions of the hole localized in a spherical NC, those for a cube-shaped NC can not be written in a simple analytical form.  We find the hole energy spectrum and wave functions  numerically using our previously developed method  (see Ref. [\onlinecite{Semina2021}]). The wave functions are found as the expansion on the basis of hole eigenfunctions in the parabolic band model in the same NC. Then the matrix elements for the kinetic energy, as well as the magnetic field contributions, are found and the eigenstates and eigenenergies are obtained by diagonalizing the resulting matrix-form Hamiltonian.  Here, we also expand our method for an arbitrary oriented magnetic field.

\begin{figure}
\includegraphics[width=0.99\columnwidth]{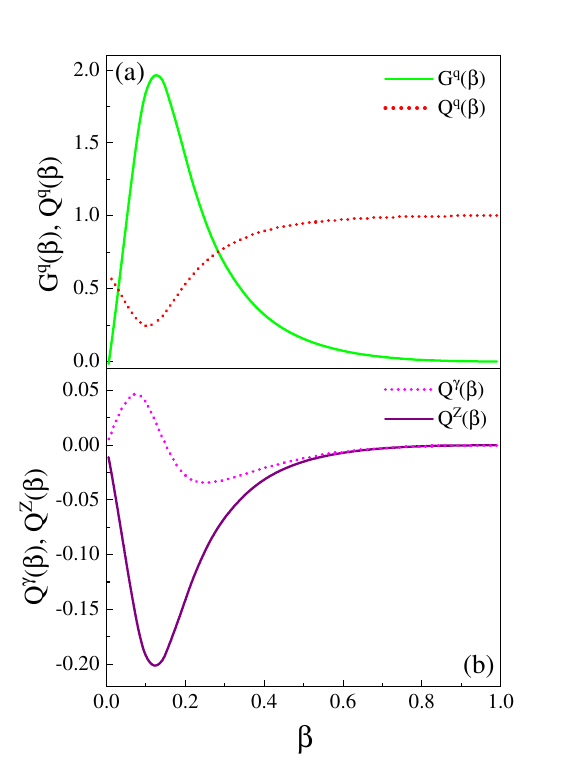}
   \caption{Dependencies of functions (a) $G^q(\beta)$ and $ Q^q(\beta)$; (b) $Q^{\gamma}(\beta)$ and $Q^Z(\beta)$ on the light to heavy hole effective mass ratio $\beta$ for cube-shaped NCs, calculated numerically.  }
 \label{fig:4}
\end{figure}

The cubic shape of the NC leads to additional contributions to $g_h$ and $Q^{\text{eff}}$ as compared with spherical NCs, Eqs. \eqref{Gelmontgen} and \eqref{Qq}.  Phenomenologicaly,  one can write
\begin{equation}\label{gh_cube}
g_h=I(\beta)\gamma_1+2S(\beta)\varkappa+G^q(\beta)q,
\end{equation} 
\begin{equation}\label{Q_eff_cube}
Q^{\text{eff}}=Q^{\gamma}(\beta)\gamma_1+Q^Z(\beta)\varkappa+Q^q(\beta)q.
\end{equation} 
Here, functions  $I(\beta)$,  $S(\beta)$, $G^q(\beta)$,  $Q^{\gamma}(\beta)$, $Q^Z(\beta)$ and $Q^q(\beta)$  depend only on the light to heavy hole effective mass ratio $\beta$, and on the type and shape of the NC potential. Function $G^q(\beta) \propto Q_2$ describes the contribution of the cubically-symmetric term in the Zeeman Hamiltonian \eqref{HZ} to the isotropic part of the effective Zeeman Hamiltonian \eqref{effect_Z}. Importantly, here $Q_2 \ne 0$ only due to the cubic symmetry of the hole wave functions inside a NC. In addition, the cubic symmetry  of the wave functions induces two absent in spherical NCs contributions into $Q^{\text{eff}}$ term proportional to the parameters of the isotropic Luttinger Hamiltonian and described by the functions $Q^{\gamma}(\beta)$ and $Q^Z(\beta)$. 
Expressions \eqref{gh_cube} and \eqref{Q_eff_cube} are reduced to Eqs. \eqref{Gelmontgen} and  \eqref{Qq} for spherical NCs, if one take $G^q(\beta)\equiv Q^{\gamma}(\beta)\equiv Q^Z(\beta)\equiv 0$.  

For cube-shaped NCs, all functions  from Eqs. \eqref{gh_cube} and \eqref{Q_eff_cube} are non-zero and have to be calculated numerically. The dependencies $I(\beta)$ and  $S(\beta)$ are shown in Fig. \ref{fig:1}, $G^q(\beta)$, $Q^q(\beta)$, $Q^{\gamma}(\beta)$  $Q^Z(\beta)$  are sown in Fig. \ref{fig:4}. Interestingly, for cube-shaped NCs, unlike spherical ones, $Q^q(\beta)\neq S(\beta)$. This fact is the consequence of difference between the structure of hole wave function in spherical and cube-shaped NCs.
From Fig. \ref{fig:4}(b) one can see, that in cube-shaped NCs  contributions to $Q^{\text{eff}}$ from the orbital magnetic field contribution $\widehat{H}_B$ and from the isotropic part of the Zeeman Hamiltonian can be quite large for typical $\beta\approx0.2$ in the semiconductors we are interested in (see Table \ref{Table_param}). It results in substantial hole Zeeman splitting anisotropy, even in the absence of  valence band warping.  The parameter $Q^q(\beta)$ in cube-shaped NCs, as in spherical NCs, is less than unity and leads to only a small contribution from the crystallographic Zeeman Hamiltonian $\propto q$ to $Q^{\text{eff}}$. A larger value of $G^q(\beta)$ also leads to small renormalization of the isotropic $g_h$ part of the hole $g$ factor due to a small value of $q$. \newline

\subsection{The effect of valence band warping}  \label{sec_wap}

The spherical approximation for the Luttinger Hamiltonian, Eq. \eqref{lutt_sph}, is widely used and has proved itself  effective in calculating  the energy of the hole ground state. This approximation gives a correct hole ground state energy in any spherically-symmetric potential in first order perturbation theory, while the second order corrections are also usually small as in typical semiconductors $\gamma_3-\gamma_2\ll \gamma_3+\gamma_2$ \cite{Efros1998}.
On the other hand, the effect of the valence band warping, $\gamma_2\neq\gamma_3$,  on the hole $g$ factor is expected to be stronger due to non-zero first order corrections to the hole wave functions. Indeed, the renormalization of the hole $g$ factor in nanostructures is controlled to a large extent by the mixing of hole states with different momentum projections on the magnetic field \cite{Semina2021}. This mixing is modified strongly by the kinetic energy term $\propto \alpha$, Eq. \eqref{lutt_matrix_warp},  with  $\alpha=(\gamma_3-\gamma_2)/\gamma$ being the parameter characterizing  valence band warping. This effect is qualitatively similar to the effect of the cubic shape of NCs in the case $\gamma_2=\gamma_3$, considered in previous section.

Here, we discuss the effect of  valence band warping on the hole Zeeman splitting in an external magnetic field. Unlike the cubically-symmetric contribution to the Zeeman Hamiltonian (\ref{HZ}),  valence band warping cannot be taken into account by  first order perturbation theory with wave functions from  Eq. \eqref{Gelmont_functions} as  good zero order approximation functions. The correct functions must account for  valence band warping even in zero magnetic field \cite{Gelmont1991,Merkulov1994w,Malyshev1996,Malyshev1997}, and these functions cannot be found analytically. The full analysis of the effect of valence band warping on the hole effective $g$ factor for arbitrarily $\beta$ is beyond the scope of this work. Here, we  confine ourselves to the numerical analysis of the warping effect in a number of semiconductors.

To obtain the Zeeman splitting of the hole ground state in spherical parabolic  potential and cube-shaped NCs, we calculate first hole wave functions in zero magnetic field using our numerical methods  developed for spherical approximation of the Luttinger Hamiltonian and  modified appropriately to take valence band warping into account \cite{Semina2021,Semina2016}. As we use the full basis (accurate within its finite size in real calculation), obtained results are valid also when the non-spherical kinetic energy term of the hole Hamiltonian is included. We develop additionally the numerical method for calculating hole wave functions in spherical NCs with box-like potential  accounting for the valence band warping as described in Appendix B. Importantly, even in spherical NCs the account of the valence band warping results into cubical anisotropy of the hole spatial distribution. Then, the Zeeman splitting in magnetic field is found using  and $\hat H_Z + \hat H_B$ as the perturbation, where $\hat H_B$ includes both isotropic and cubically-symmetric therms $\propto \alpha$ (see Appendix A).

\begin{figure}
\includegraphics[width=0.75\columnwidth]{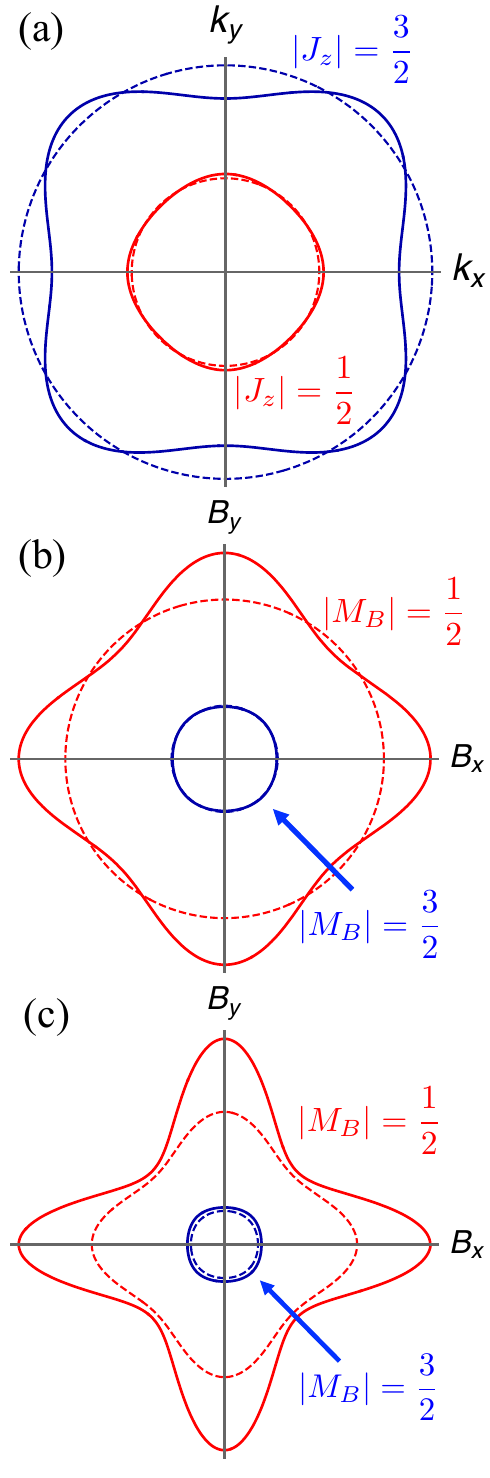}
   \caption{(a) Constant-energy curves of bulk heavy, $J_z$=3/2 (blue color), and light holes, $J_z$=1/2 (red color), in $xy$ plane, (b) and (c) Constant-Zeeman-splitting curves  for  hole states with $|M_B|=3/2$ blue color and $|M_B|=1/2$ red color calculated for spherical  and cube-shaped NCs with box-like potential, correspondingly.  Calculations were made for CdSe parameters $\gamma_1=2.52$, $\gamma_2=0.65$, $\gamma_3=0.95$ \cite{Fu1998n2}, solid lines correspond to accounting for valence band warping, dashed lines correspond to spherical approximation of the Luttinger Hamiltonian with $\gamma_2=\gamma_3=\gamma=0.83$.} 
 \label{E_warp}
\end{figure}

The valence band warping leads to non-spherical hole constant-energy surfaces in bulk semiconductor. As an example in Fig. \ref{E_warp}(a) we show  the constant-energy  curves  of bulk heavy and light holes in $xy$ plane in zero magnetic field, calculated for CdSe parameters $\gamma_1=2.52$, $\gamma_2=0.65$, $\gamma_3=0.95$ (corresponding to $\gamma=0.83$) \cite{Fu1998n2}, solid lines correspond to accounting for the valence band warping, dashed lines correspond to the spherical approximation of the Luttinger Hamiltonian. In Fig. \ref{E_warp}(a) one can clearly see cubic symmetry of constant-energy curves and more pronounced anisotropy for heavy holes if $\gamma_2\neq\gamma_3$.  In Fig. \ref{E_warp} (b) and Fig. \ref{E_warp} (c)  we show the  constant-Zeeman-splitting curves in external magnetic field ${\bm B} =(B_x,B_y,0)$ for states with $|M_B|=3/2$ and $|M_B|=1/2$ calculated for spherical and cube-shaped NCs, respectively,  with the same Luttinger parameters. In contrast to the Fig.~\ref{E_warp}(a), the anisotropy is more pronounced for holes with $|M_B|=1/2$.  Due to the equivalence of all cubic axes, the picture would be the same for any plane containing two cubic axes. It is noteworthy, that in the presence of the uniaxial splitting $\Delta_{\rm an}$ of the localized hole states  with $|M|=3/2$ and $|M|=1/2$ (see Fig. \ref{g_angle}(a,b)), which allows one to neglect their mixing, the transverse $g$ factor in the plane perpendicular to the anisotropy axis becomes isotropic, even with accounting for valence band warping. The only anisotropy in that case would be with respect to the angle with the $z$-axis (transverse or longitudinal $g$ factor).

\begin{figure*}
\includegraphics[width=0.99\textwidth]{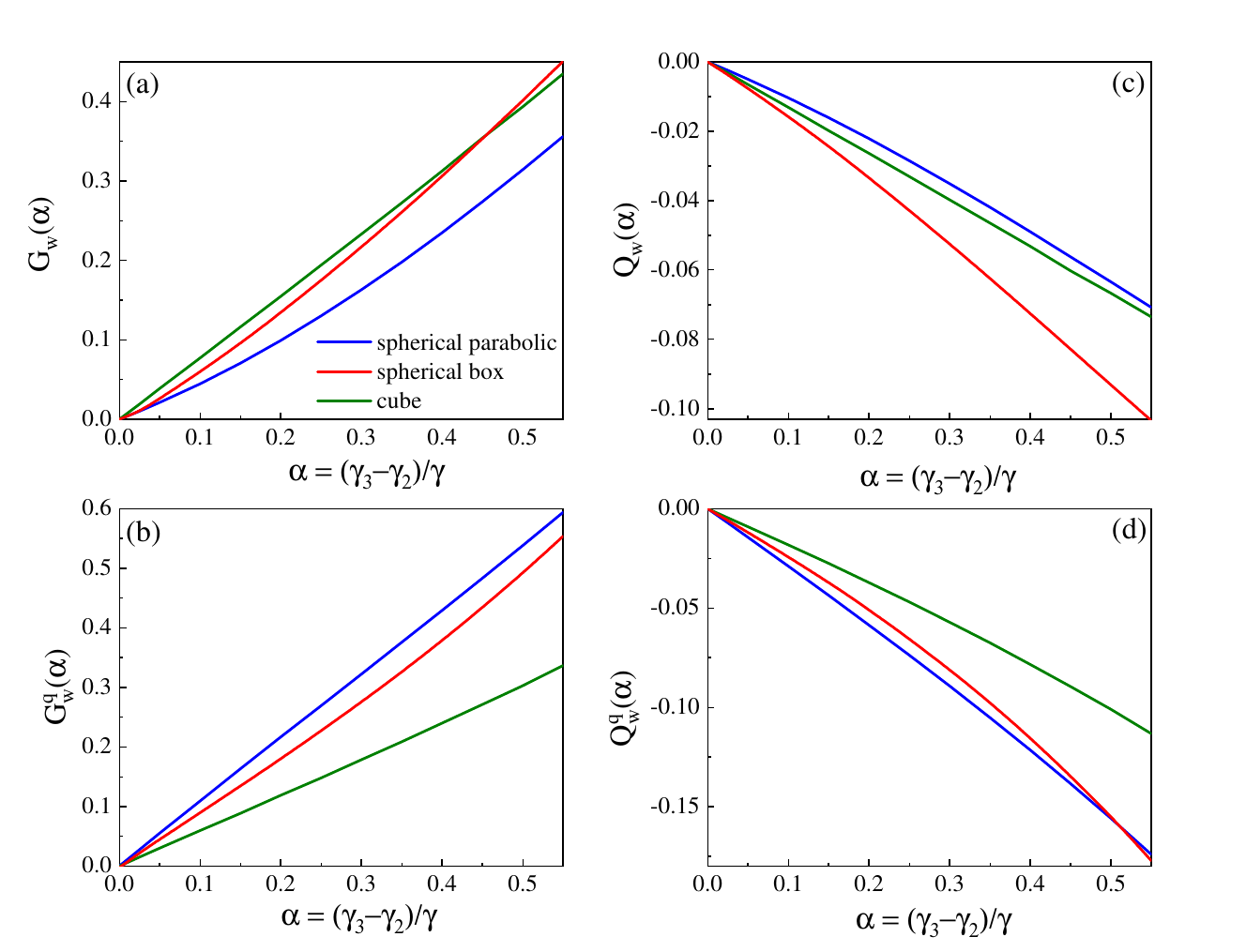}
   \caption{Functions (a) $G_{\text w}(\alpha)$, (b) $G^q_{\text w}(\alpha)$, (c) $Q_{\text w}(\alpha)$  and (d) $Q^q_{\text w}(\alpha)$  calculated numerically for CdSe valence parameters  \cite{Fu1998n2}. }
 \label{G_warp}
\end{figure*}

Let us proceed to a quantitative analysis of the hole $g$ factor anisotropy. The $g$ factor components $g_{\text{h}}$ \eqref{gh_cube} and $Q^{\text{eff}}$ \eqref{Q_eff_cube} now also depend on the valence band warping parameter $\alpha$. For illustrative purposes, we numerically calculated  those dependencies for  fixed valence band parameters for zinc-blende CdSe: $\gamma_1=2.52, ~\gamma=0.83$ \cite{Fu1998n2}, changing $\gamma_2$ and $\gamma_3$ in such  a way, that value of $\gamma=(2\gamma_2+3\gamma_3)/5=0.83$ and $\beta=2$ remain constant. Simultaneously, the value of $\varkappa=-0.12$, determined from Eq. \eqref{kappa}  is also constant.
As we  are interested now only in dependences of $g_h$ and $Q^{\rm eff}$ on $\alpha$, it is convenient to write them  as:
\begin{equation}\label{gh_cube_warp}
g_h(\alpha)=G_0+G_{\text w}(\alpha)+\left(G_0^q+G^q_{\text w}(\alpha)\right) q,
\end{equation} 
\begin{equation}\label{Q_eff_cube_warp}
Q^{\text{eff}}(\alpha)=Q_0+Q_{\text w}(\alpha)+\left(Q_0^q+Q^q_{\text w}(\alpha)\right) q . 
\end{equation} 
Here $G_0$, $G_0^q$,  $Q_0$ and $Q_0^q$ describe $g_{\text{h}}$ and $Q^{\text{eff}}$ calculated after Eqs. \eqref{gh_cube} and \eqref{Q_eff_cube} for $\alpha=0$ and fixed values of $\beta$ and $\varkappa$. 
Accounting for $\alpha \ne 0$ leads to cubically-symmetric shape of hole wave function in zero magnetic field, and, consequently, to additional contributions to  $Q_2$ tensor component even in spherical NCs. This results in new contributions to both $g_h$, described by the functions $G_{\text w}(\alpha)$ and $G^q_{\text w}(\alpha)$,  and $Q^{\rm eff}$, described by the functions $Q_{\text w}(\alpha)$ and $Q^q_{\text w}(\alpha)$, respectively.  

The numerically calculated dependencies of $G_{\text w}(\alpha)$, $G^q_{\text w}(\alpha)$, $Q_{\text w}(\alpha)$ and $Q^q_{\text w}(\alpha)$  for chosen CdSe parameters for spherical NCs with parabolic and box-like potentials, and cube-shaped NCs are shown in Fig. \ref{G_warp}. For $\gamma_2=0.65$ and $\gamma_3=0.95$ \cite{Fu1998n2} (parametrization ``1'' in Table \ref{Table_param}) we get $\alpha\approx 0.36$. Corrections $G_{\text w}(\alpha)$ and $Q_{\text w}(\alpha)$ are quite  large at $\alpha\approx 0.36$, leading to noticeable renormalization of  $g_{\text{h}}$ and $Q^{\text{eff}}$  due to valence band warping. 
The dependence of hole $g$ factors on the magnetic field direction in spherical NCs with parabolic confining potentials corresponding to parameters of CdSe, parametrization ''1'' in Table \ref{Table_param},  is shown in Fig. \ref{g_angle}(f).  One can see  a strong difference to the spherical approximation for light holes and a rather small difference for heavy holes. In Figs. \ref{g_angle}(d,e), the dependencies of hole $g$ factors on the magnetic field direction are shown for the case of split states of heavy and light holes. Here the difference is smaller than in the spherical case.    
\begin{table}
\begin{center}
\caption{Valence band parameters used for the calculation of the hole Zeeman splitting. } 
\label{Table_param}
   \begin{tabular}{| l | l | l | l | l |  l | l| l|l|l|   }
   \hline
  No.&  Material & $\gamma_1$ & $\gamma_2$&$\gamma_3$&$\beta$&$\varkappa$&$\alpha$& $q$ &Refs.$^*$\\ \hline
  1&    zb-CdSe& 2.52 & 0.65& 0.95&0.2&-0.12&0.36&- &\cite{Fu1998n2} \\ \hline
 2&   zb-ZnSe& 3.94 & 1.00&1.52&0.2&0.21&0.4&- &\cite{Adachi2004} \\ \hline
   3&       CdTe& 4.14 & 1.09&1.62&0.19&0.3&0.38 &-&\cite{Friedrich1994} \\ \hline
4&  GaAs& 6.79 &1.92&2.68&0.18&1.03&0.32&0.017 &\cite{Molenkamp1988,Marie1999} \\ \hline
 \end{tabular}
\end{center}
$^*$ References are given for the $\gamma_{1}$, $\gamma_{2}$, $\gamma_{3}$ and $q$ Luttinger parameters, and  we use relations $\beta=(\gamma_1-2\gamma)/(\gamma_1+2\gamma)$ and $\varkappa\approx 2/3+5\gamma/3-\gamma_1/3$   with $\gamma=(2\gamma_2+3\gamma_3)/5$ and $\alpha=(\gamma_3 - \gamma_2)/\gamma$.
\end{table}
\begin{table*}
\begin{center}
\caption{Parameters $g^{||}_{3/2}$ , $g^{||}_{1/2}$, $g^{\perp}_{3/2}$, $g^{\perp}_{1/2}$, $g_{\text{h}}$ and $Q^{\text{eff}}$ calculated for semiconductor NCs with different shape and type of localizing potential, numbers brackets corresponds to spherical approximation of Luttinger Hamiltonian.  Material parameters used for calculations are given in Table I.  }
\label{Table_g_sph}
   \begin{tabular}{| l | l | l | l | l |  l | l | l |l|   }
   \hline
   \multicolumn{8}{| c |}{Spherical NCs with parabolic potential}\\ \hline
  No&  Material&  $g^{||}_{3/2}$ &$g^{||}_{1/2}$&$g^{\perp}_{3/2}$&$g^{\perp}_{1/2}$&$g_{\text{h}}$ ($g_{\text{h}}^{\rm sph}$) &$Q^{\text{eff}}$\\ \hline
  1&    zb-CdSe&     -0.97 (-0.98)&-0.8(-0.98)&-0.04(0)&-1.99 (-1.96)&-0.777(-0.98)&-0.04(0) \\ \hline
 2&   zb-ZnSe&   -0.85(-0.86)&-0.49(-0.86)&-0.09(0)&-1.79(-1.72)&-0.444(-0.86)&-0.09(0) \\ \hline
   3&       CdTe&  -0.8(-0.82) &-0.45(-0.82)&-0.09 (0)&-1.7(-1.64)&-0.4(-0.82)&-0.09(0) \\ \hline
4&  GaAs, $q=0$& -0.58 (-0.61)&-0.006 (-0.61)&-0.14 (0)&-1.3 (-1.22)&0.067(-0.61)&-0.14(0) \\ \hline
5&  GaAs, $q=0.017$& -0.52 (-0.61)&0.007 (-0.61)&-0.13 (0)&-1.16 (-1.22)&0.07(-0.61)&-0.13(0) \\ \hline
   \multicolumn{8}{| c |}{Spherical NCs with box-like potential}\\ \hline
    No.&  Material&  $g^{||}_{3/2}$ &$g^{||}_{1/2}$&$g^{\perp}_{3/2}$&$g^{\perp}_{1/2}$&$g_{\text{h}}$ ($g_{\text{h}}^{\rm sph}$)&$Q^{\text{eff}}$\\ \hline
  1&    zb-CdSe&     -1.08(-1.06)&-0.82(-1.06)&-0.064 (0)&-2.21 (-2.12)&-0.78 (-1.06)&-0.064 (0) \\ \hline
 2&   zb-ZnSe&   -1.02 (-1)&-0.55 (-1)&-0.12(0)&-2.17 (-2)&-0.5(-1.03)&-0.12(0) \\ \hline
   3&       CdTe&  -1.01 (-1.03) &-0.54 (-1.01)&-0.12 (0)&-2.16 (-2.2)&-0.48(-1.01)&-0.12(0) \\ \hline
4&  GaAs, $q=0$& -1.08 (-1.06)&-0.36 (-1.06)&-0.18(0)&-2.34 (-2.12)&-0.27(-1.06)&-0.18(0) \\
\hline
5&  GaAs, $q=0.017$& -1.05 (-1.06)&-0.355 (-1.06)&-0.173(0)&-2.26 (-2.12)&-0.26(-1.06)&-0.17(0) \\ \hline
\multicolumn{8}{| c |}{Cube-Shaped NCs}\\ \hline
 No.&  Material&  $g^{||}_{3/2}$ &$g^{||}_{1/2}$&$g^{\perp}_{3/2}$&$g^{\perp}_{1/2}$&$g_{\text{h}}$&$Q^{\text{eff}}$\\ \hline
  1&    zb-CdSe&   -0.87 (-0.96)&-0.47 (-0.73)&-0.1 (-0.058)&-1.84 (-1.98)&-0.42(-0.7)&-0.1(-0.058)\\ \hline
 2&   zb-ZnSe&  -0.76 (-0.87)&-0.17 (-0.3)&-0.15 (-0.14)&-1.68 (-1.88)&-0.09(-0.24)&-0.15(-0.14) \\ \hline
   3&       CdTe& -0.75 (-0.86)&0.24 (-0.26)&-0.25 (-0.15)&-1.76 (-1.86)&0.37(-0.19)&-0.25(-0.148) \\ \hline
4&  GaAs, $q=0$& 0.58 (-0.72)&1.3 (0.42)&-0.18(-0.29)&1 (-1.8)&1.4(0.58)&-0.18(-0.295) \\ \hline
5&  GaAs, $q=0.017$ & 0.5 (-0.72)&1.33 (0.42)&-0.2(-0.29)&0.8 (-1.8)&1.44(0.58)&-0.2(-0.295) \\ \hline
 \end{tabular}
\end{center}
 \end{table*}
 
Parameters $g^{||}_{3/2}$ , $g^{||}_{1/2}$, $g^{\perp}_{3/2}$ and $g^{||}_{1/2}$ in spherical NCs with parabolic and box-like infinite potentials and cube-shaped NCs with box-like infinite potential, calculated for several semiconductors (for parameters see Table \ref{Table_param}) are shown in Table \ref{Table_g_sph}. Numbers in brackets correspond to the spherical approximation of the Luttiger Hamiltonian.  As we know the value of $q$ only for GaAs, all calculations were made for $q\equiv 0$. Corresponding values of $g_h$ and $Q^{\text{eff}}$ are also shown  in Table \ref{Table_g_sph}. The contribution to    $Q^{\text{eff}}$  induced by valence band warping is quite substantial and  comparable to the contribution from the cubic shape of  NC. A similar effect was reported for acceptors  [\onlinecite{Malyshev2000}] and  disk-like quantum dots  [\onlinecite{Trifonov2021}], where the quite large cubically-symmetric contribution to hole Zeeman splitting coming from  valence band warping was reported. Note, that depending on semiconductor,  contributions to $Q^{\text{eff}}$ from cubic shape of NC and valence band warping in cube-shaped NCs could be of the same or opposite signs.  As one can see   comparing Fig. \ref{E_warp}(b) and Fig. \ref{E_warp}(c), the cubic anisotropy of the light hole Zeeman splitting is pronounced already in spherical NCs made of zb-CdSe and is enhanced by factor 2 in cube-shaped NCs. Although, corrections to $g_h$ and $Q^{\text{eff}}$ are noticeable, for heavy hole they numerically almost compensate each other resulting in much weaker anisotropy, see corresponding curves in   Fig. \ref{E_warp}(b) and Fig. \ref{E_warp}(c).

As  mentioned above, we made all our estimations of the effect of valence band warping on hole $g$ factors for $q\equiv 0$ because only the value for GaAs is known from the literature. Here, we demonstrate the smallness effect of $q\neq 0$. We made a calculation taking into account $q\neq 0$ for GaAs. The results are also shown in Table \ref{Table_g_sph} in lines with label $5$. One can see a small difference, as compared to lines with label $4$,  corresponding to the same set of Luttinger parameters and $q=0$.

\section{Possible experimental manifestations of the hole $g$ factor cubic anisotropy}

In this section, we discuss possible experimental consequences of the non-zero value of $Q^{\rm eff}$.
Let us firstly discuss manifestations of the hole $g$ factor anisotropy in the case of colloidal nanostructures with split states of light and heavy holes (see Figure \ref{g_angle}(a)), such as spheroidal or cuboidal nanocrystals, where splitting can reach tens of meV, or nanoplatelets with a splitting up to $300$~meV \cite{Ithurria2011nm}. In such structures, the value of the transverse $g$ factor of the heavy hole, $g^{\perp}_{3/2}$, is determined only by the cubically-symmetric contribution $Q^{\rm eff}$ to the effective Zeeman Hamiltonian. Due to a large splitting of the light and heavy hole states in these structures, a fast hole spin relaxation, usually observed in bulk semiconductors, should be suppressed. It opens the possibility for detection of the $Q^{\rm eff}$ related contribution from studies of the hole spin precession, as  was done for epitaxial quantum wells \cite{Marie1999,Syperek2007, Kugler2009}. 

Another possible  manifestation concerns the degree of circular polarization (DCP) of photoluminescence in a magnetic field for ensembles of neutral colloidal nanocrystals or nanoplatelets. At low temperatures, the photoluminescence properties of these nanostructures are determined by emission of dark excitons with the projection of the total angular momentum of an electron and a hole $F_z=2$ on the anisotropy axis \cite{efros1996} . It was shown theoretically in \cite{JohnstonHalperin2001} that  DCP should reach $75 \%$ for an ensemble of randomly oriented NCs in high magnetic fields. However, this upper limit was never observed experimentally. It was proposed previously that a low saturation value of DCP is caused by  nonradiative recombination of dark excitons \cite{JohnstonHalperin2001}, or by linearly polarized phonon-assisted emission of dark excitons \cite{qiang2020}. 

\begin{figure}
\includegraphics[width=0.75\columnwidth]{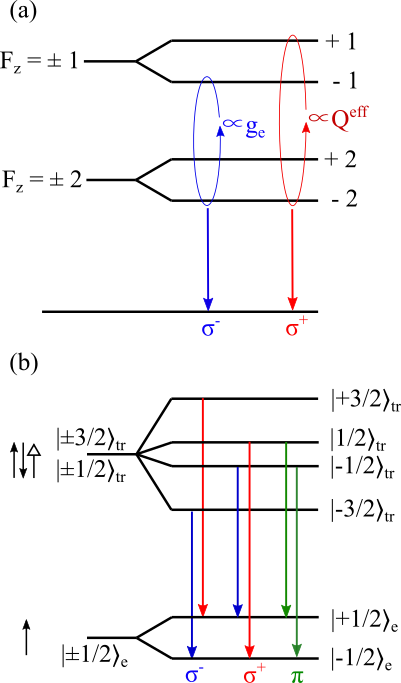}
   \caption{(a) Schematic of the Zeeman splitting of the dark $F_z=\pm 2$ and bright $F_z=\pm1$ exciton states in NCs with a large anisotropic splitting $\Delta_{\rm an}$. Arrows show admixture of bright exciton states to the lowest dark state $F_z=-2$ resulting in its emission with $\sigma^+$ or $\sigma^-$ polarization. (b) Optical transitions for a negatively charged trion with magnetic field applied at angle $\theta=0$ with respect to the crystal $z$-axis. Blue, red, and green arrows show transitions with $\sigma^-$, $\sigma^+$, and $\pi$ polarization, respectively. Black and white arrows show spin of the electron and hole, respectively.} 
 \label{figtrion}
\end{figure}

Our analysis shows that a nonzero transverse $g$ factor of the heavy hole, $g^{\perp}_{3/2}\propto Q^{\rm eff}$, also can affect the saturation value of DCP. In \cite{JohnstonHalperin2001}, it was assumed that dark exciton states with $F_z=\pm2$ gain  oscillator strength in a transverse magnetic field due to mixing of electron spin states $\pm1/2$. Under this condition, the bright exciton state with $F_z=+1(-1)$ admixes to the state $F_z=+2(-2)$, respectively. The dark exciton states $F_z=\pm2$ inherit optical properties of the bright exciton states and emit $\sigma^{\pm}$ polarized light along the anisotropy axis of a nanocrystal, respectively (see Fig. \ref{figtrion}(a)). The nonzero component $g^{\perp}_{3/2}$ allows admixture of the bright exciton state $F_z=+1(-1)$ to the dark exciton state $F_z=-2(+2)$. It results in opposite circular polarization of photons emitted by the dark exciton states and in a decrease of the DCP saturation value. The decreasing factor depends on the angle $\theta$ between the anisotropy axis of a NC and magnetic field direction. For a given angle $\theta$, this factor equals $(a_{-1}^2-a_{+1}^2)/(a_{-1}^2+a_{+1}^2)$, where $a_{+1/-1}$ are amplitudes of the $F_z=\pm1$ exciton states in the wavefunction of the $F_z=-2$ dark exciton state, which is usually the lowest in energy (i.e. predominantly populated) exciton state. Amplitudes $a_{-1}$ and $a_{+1}$ are proportional to electron $g$ factor, $g_e$, and parameter $Q^{\rm eff}$, respectively. It means that depending on the ratio $g_e/Q^{\rm eff}$, the value of the dark exciton DCP can vary in a wide range.

Now, let us consider cube-shaped NCs, where a large anisotropy of the hole $g$ factor is expected. In these NCs, the ground hole state is four-fold degenerate in the absence of a magnetic field. Thus, one cannot expect experimental observation of $Q^{\rm eff}$ related effects using time-resolved spectroscopy due to a fast hole spin decoherence. We suppose that observation of these effects can be realised by photoluminescence spectroscopy of single cube-shaped NCs in an applied magnetic field. 
For this purpose, the best candidate is the negatively charged trion (a hole plus two electrons in a singlet state), since its spin splitting is determined solely by the hole $g$ factor and there is no exchange interaction  due to the singlet electrons configuration. 

In our considerations, we assume that a cube-shaped NC stands on a substrate, and its edges are directed along the axes of the laboratory coordinate system, as  shown in Fig.~\ref{g_angle}(c). A magnetic field is applied in the $xz$ plane at an arbitrary angle $\theta$. The photoluminescence signal is determined by photons emitted along the $z$ axis. The excited state of the system is the negative trion. Its spin splitting in an applied magnetic field is determined by the spin splitting of hole spin states.  After the radiative recombination of the trion, there remains one electron with the spin projection up or down in the magnetic field direction.
For a tilted magnetic field, the excited and ground eigenstates of the system are superpositions of the hole and electron basis states along the $z$-axis, respectively. It results in an angular dependence of the energy and oscillator strength of trion emission lines.  
A schematic of the energy states of the system and possible allowed transition for magnetic field directed in the Faraday geometry (along the $z$-axis of the laboratory frame) is shown in Figure \ref{figtrion}(b).  Details of the calculation can be found in Appendix C.

\begin{figure*}[!ht] \centering \includegraphics[width=\textwidth]{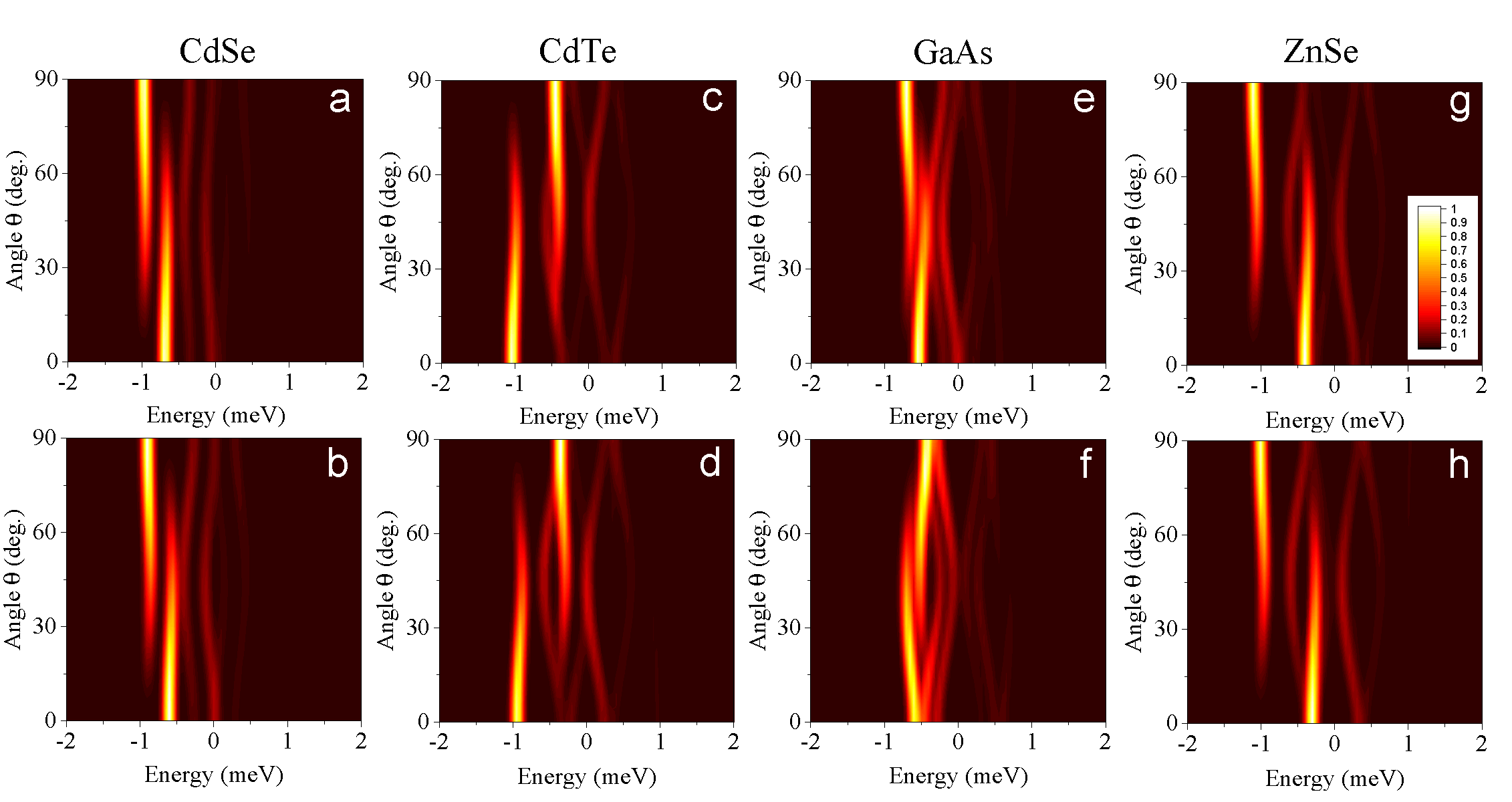} \caption{Calculated spectra of the negative trion at $T = 4$~K and $B=10$~T in cube-shaped NCs of $15$~nm size for (a,b) CdSe (c,d) CdTe (e,f) GaAs (g,h) ZnSe. The upper row is calculated for the spherical approximation of the Luttinger Hamiltonian with $\gamma_2=\gamma_3$. The bottom row is calculated for the full Luttinger Hamiltonian with $\gamma_2\neq\gamma_3$.  } \label{fig:trionlines} 
\end{figure*}

We performed calculations for cube-shaped NCs made of CdSe, CdTe, GaAs and ZnSe with edge size of $15$~nm, according to a typical size from \cite{Lv2022}. In our model, the only size-dependent parameter is the electron $g$ factor. Its size dependence can be found within the $\bm {kp}$-method \cite{Semina2021}. For the NC size of $15$~nm, the calculated electron $g$ factor equals:
$g_e({\rm CdSe})=0.5,~g_e({\rm CdTe})=-1,~g_e({\rm GaAs})=0.3,~g_e({\rm ZnSe})=1.2$. 
The upper row in Figure \ref{fig:trionlines} corresponds to the spherical approximation of the Luttinger Hamiltonian with $\gamma_2=\gamma_3$. The bottom row in Figure \ref{fig:trionlines} corresponds to the full Luttinger Hamiltonian with $\gamma_2\neq\gamma_3$. One can see that even in the spherical approximation of the Luttinger Hamiltonian,  the NC cubic shape affects the angular dependence of the energies of the trion emission lines. Accounting for  valence band warping changes the values of the trion lines splitting but does not strongly modify their angular dependencies.  We note, that in the absence of valence band warping and the NC cubic shape, the energies of the trion emission lines do not depend on the direction of the applied magnetic field. Thus, experimental observation of the angular dependencies shown in Fig.\ref{fig:trionlines} should give evidence of the hole $g$ factor anisotropy.  

It is worth noting that the cubic symmetry contribution to the Zeeman splitting cannot arise in perovskites or CuCl nanocrystals even of cubic shape. In these materials, the conduction and valence bands are characterized by a total angular momentum $j$ of electron or hole equal to 1/2. In perovskites,  the anisotropy of charge carriers $g$ factors appears only if the symmetry of the structure, of crystal lattice or of NCs shape is lower than cubic \cite{Kirstein2022}.

\section{Conclusion}

We demonstrated that the Zeeman splitting of hole states inherits  both the shape and lattice symmetry of nanocrystals made of zinc-blende semiconductors. We show that the cubic shape of a nanocrystal and cubic symmetry of the zinc-blende crystal lattice make comparable contributions to the hole Zeeman splitting.  In both cases, effects arise due to the cubically-symmetric instead of spherically-symmetric spatial distribution of the hole wave functions and, consequently, renormalized light hole to heavy hole mixing.  The resulting hole  Zeeman splitting is non-equidistant and depends on the direction of the magnetic field with respect to the cubic axes. 
We show that both cubically-symmetric contributions are large enough and should be taken into account in analysis of magneto-optical studies of spin splitting of hole states. Possible experimental manifestations of the predicted cubic anisotropy of hole Zeeman splitting in semiconductor nanocrystals are discussed.

\section*{Acknowledgments}
We thank D.R. Yakovlev for valuable discussions. \newline
This work was funded by the Russian Science Foundation (Grant No. 23-12-00300).

\appendix
\setcounter{equation}{0}
\setcounter{figure}{0}
\setcounter{table}{0}
\renewcommand{\thefigure}{A\arabic{figure}}
\renewcommand{\theequation}{A\arabic{equation}}
\renewcommand{\thetable}{A\arabic{table}}

\section{Matrix form of hole Hamiltonians}
 \label{AA}

We present the Luttinger Hamltonian describing hole kinetic energy in two terms: $H_L = H_L^{\rm sph}+\alpha H_L^{\alpha}$, where the fist one represents the spherically symmetric part and the second describes the valence band warping with its cheracterizing parameter $\alpha=(\gamma_2-\gamma_3)/\gamma$.
In the standard basis of the topmost $\Gamma_8$ valence subband $u_\mu$ ($\mu=+3/2,+1/2,-1/2,-3/2$)\cite{Bir1974book,Ivchenko2005book}  Hamiltonians $H_L^{\rm sph}$ and $H_L^{\alpha}$ in matrix form are:
\begin{multline}\label{lutt_matrix_sph}
\hat{H}_L^{\rm sph}=\frac{\hbar^2}{2m_0}\left(
\begin{array}{cccc}
 P+Q & H & I &
   0 \\
 H^\dag &  P-Q & 0 & I \\
 I^\dag & 0 &  P-Q & -H \\
 0 & I^\dag & -H^\dag &  P+Q \\
\end{array}
\right),\end{multline}
$$P=\gamma_1(k_x^2+k_y^2+k_z^2),
Q=\gamma(k_x^2+k_y^2-2k_z^2),$$
$$H=-2 \sqrt{3} \gamma k_z (k_x-\mathrm i k_y), I=-\sqrt{3} \gamma (k_x-\mathrm i k_y)^2.$$
\begin{multline}\label{lutt_matrix_warp}
\hat{H}_L^{\rm \alpha}= \frac{\hbar^2}{2m_0}\left(
\begin{array}{cccc}
  Q^{\alpha} & H^{\alpha} &I^{\alpha} & 0 \\
 (H^{\alpha})^\dag & -Q^{\alpha} & 0 & I^{\alpha} \\
(I^{\alpha})^\dag & 0 & -Q^{\alpha} & -H^{\alpha} \\
 0 & (I^{\alpha})^\dag & -(H^{\alpha})^\dag & Q^{\alpha} \\
\end{array}
\right),\end{multline}
$$Q^{\alpha}=-\frac{3}{5}\gamma(k_x^2+k_y^2-2k_z^2),
H^{\alpha}=-\frac{4 \sqrt{3}}{5} \gamma k_z (k_x-{\mathrm i} k_y),$$
$$I^{\alpha}= \frac{\sqrt{3}\gamma}{5}\left(3k_x^2+4{\mathrm i}k_xk_y-3k_y^2\right).$$

It can be easily shown, that the first order correction to the hole ground state ($S_{3/2}$) energy in spherical NCs coming from  $\hat{H}_L^{\rm \alpha}$ calculated on the wave functions \eqref{Gelmont_functions} is zero.  The first order correction for cube-shaped NCs is already non-zero due to wave function structure. 
The Zeeman  Hamiltonian \eqref{HZ} in matrix form is: 
\begin{widetext}
\small{\begin{equation}\label{Hzmatrix}
\hat{H}_z=\mu_B\left(
\begin{array}{cccc}
 - (3 \varkappa +\frac{27}{4} q)B_z & - (\sqrt{3} \kappa +\frac{7\sqrt{3}}{4} q)B_{-} & 0 & -\frac{3}{2} q B_{+} \\
 -(\sqrt{3} \kappa +\frac{7\sqrt{3}}{4} q) & -
    (\varkappa +\frac{1}{4}q)B_z & - (2 \varkappa +5 q)B_{-} & 0 \\
 0 & - (2 \varkappa +5 q)B_{+} & ( \varkappa +\frac{1}{4}q)B_z &
   - (\sqrt{3} \kappa +\frac{7\sqrt{3}}{4} q)B_{-} \\
 -\frac{3}{2} q B_{-} & 0 & -   (\sqrt{3} \kappa +\frac{7\sqrt{3}}{4} q)B_{+} &  (3 \varkappa +\frac{27}{4} q) B_z\\
\end{array}
\right),
\end{equation}}
\normalsize \end{widetext}
where $B_{\pm}=B_x\pm \mathrm i B_y$.
The explicit form of orbital contribution  $H_B=H_B^{\rm sph}+ \alpha H_B^{\alpha}$ neglecting quadratic on magnetic field terms of the isotropic part, $H_B^{\rm sph}$, is:
\small{\begin{equation}\label{HBmatrix_sph}
\hat{H}_B^{\rm sph}=\mu_B\left(
\begin{array}{cccc}
 P_B+Q_B & -S_B & R_B & 0 \\
  -(S_B)^\dag & P_B-Q_B & 0 & R_B \\
  (R_B)^\dag & 0 & P_B-Q_B & S_B \\
  0 & (R_B)^\dag & (S_B)^\dag & P_B+Q_B\\
\end{array}
\right),\end{equation}
\begin{multline*} P_B=\gamma_1\left( (zk_y-yk_z)B_x+(xk_z-zk_x)B_y+(yk_x-xk_y)B_z\right),\\ Q_B=\gamma\left( (zk_y+2yk_z)B_x+(2xk_z+zk_x)B_y+(yk_x-xk_y)B_z\right),\\R_B= -\sqrt{3}\gamma_2 (zk_yB_x-zk_xB_y+(yk_x-xk_y)B_z)+\\+\mathrm{i}\sqrt{3}\gamma_2(zk_xB_x-zk_yB_y+(yk_y-xk_x)B_z), \\S_B=\sqrt{3}\gamma((xB_y-yB_x)(k_x-\mathrm i k_y)+\\+\mathrm i (x-\mathrm i y)k_zB_z-\mathrm i zk_z(B_x-\mathrm i B_y)).  \\
\end{multline*}
}\normalsize  
As we have shown, that to determine hole $g$ factor components  it is sufficient to calculate hole Zeeman splitting for $\bm B|| z$. Here we show  $\hat{H}_B^{\rm sph}$ components for $\bm B =(0,0,B_z)$:
\begin{multline} P_B=\gamma_1(yk_x-xk_y)B_z, Q_B=\gamma(yk_x-xk_y)B_z,\\R_B= -\sqrt{3}\gamma(\mathrm i x(k_x+\mathrm i k_y)+y(k_x-\mathrm i k_y))B_z. 
\end{multline}
Note, that even $H_B^{\rm sph}$ can contribute to the anisotropic part of the hole effective $g$ factor if the hole wave functions have cubic symmetry in zero magnetic field.
\small{\begin{equation}\label{HBmatrix_warp}
\hat{H}_B^{\alpha}=\mu_B\left(
\begin{array}{cccc}
Q_B^{\alpha} & -S_B^{\alpha} & R_B^{\alpha} & 0 \\
  -(S_B^{\alpha})^\dag & -Q_B^{\alpha} & 0 & R_B^{\alpha} \\
  (R_B^{\alpha})^\dag & 0 & -Q_B^{\alpha} & S_B^{\alpha} \\
  0 & (R_B^{\alpha})^\dag & (S_B^{\alpha})^\dag &Q_B^{\alpha}\\
\end{array}
\right),\end{equation}
\begin{multline*}Q_B^{\alpha}=-\frac{3\gamma}{5}( (zk_y+2yk_z)B_x+\\+(2xk_z+zk_x)B_y+(yk_x-xk_y)B_z),\\ R_B^{\alpha}=-\frac{\sqrt{3}\gamma}{5}  (-2 \mathrm i B_x z k_x -3 B_x z k_y +3 B_yz k_x +2 \mathrm i B_y z k_y+\\+2 \mathrm i B_z x k_x-3 B_z y k_x+3 B_z x k_y-2 \mathrm i B_z y k_y), \\S_B^{\alpha}=-\frac{2\sqrt{3}\gamma}{5}((xB_y-yB_x)(k_x-\mathrm i k_y)+\\+\mathrm i (x-\mathrm i y)k_zB_z-\mathrm i zk_z(B_x-\mathrm i B_y)).  \\
\end{multline*}
}\normalsize  
If $\bm B =(0,0,B_z)$ we have
\begin{multline}Q_B^{\alpha}=-\frac{3\gamma}{5}\left((yk_x-xk_y)B_z\right),\\ R_B^{\alpha}=-\frac{\sqrt{3}\gamma}{5}  (2 \mathrm i  x k_x-3  y k_x+3 x k_y-2 \mathrm i y k_y)B_z, \\S_B^{\alpha}=-\frac{2\mathrm i\sqrt{3}\gamma}{5}( (x-\mathrm i y)k_zB_z).  
\end{multline}
The matrix form of hole Zeeman effective Hamiltonian:
\begin{widetext}
\small{\begin{equation}\label{Hzeff}
\hat{H}_Z^{\text{eff}}=\mu_B\left(
\begin{array}{cccc}
 - \frac{3}{2}g^{||}_{3/2}B_z & \frac{\sqrt{3}}{4}(3g^{\perp}_{3/2}-g^{\perp}_{1/2})B_{-} & 0 & -\frac{3}{2} g^{\perp}_{3/2} B_{+} \\
 \frac{\sqrt{3}}{4}(3g^{\perp}_{3/2}-g^{\perp}_{1/2})B_{+} & - \frac{1}{2}g^{||}_{1/2}B_z & -\frac{1}{2}g^{\perp}_{1/2}B_{-} & 0 \\
 0 & -\frac{1}{2}g^{\perp}_{1/2}B_{+} &  \frac{1}{2}g^{||}_{1/2}B_z &
   \frac{\sqrt{3}}{4}(3g^{\perp}_{3/2}-g^{\perp}_{1/2})B_{-} \\
 -\frac{3}{2} g^{\perp}_{3/2} B_{-} & 0 & \frac{\sqrt{3}}{4}(3g^{\perp}_{3/2}-g^{\perp}_{1/2})B_{+} &  \frac{3}{2}g^{||}_{3/2}B_z\\
\end{array}
\right),
\end{equation}}
\normalsize \end{widetext}
$g_{3/2}^{||}=g_{\rm h}+\frac{9}{2}Q^{\text{eff}}$, $g_{1/2}^{||}=g_{\rm h}+\frac{1}{2}Q^{\text{eff}}$,
$g_{3/2}^{\perp}=Q^{\text{eff}}$, $ g_{1/2}^{\perp}=2g_{\rm h}+10Q^{\text{eff}}$.
 If the states with $|M|=3/2$ and $|M|=1/2$ are split, the matrix elements of Hamiltonian \eqref{Hzeff} mixing them with each other are zero and matrix Eq. \eqref{Hzeff} can be separated into two independent matrices written in the basis of Bloch states $u_\mu$ with $\mu=+3/2, -3/2$ and $\mu=+1/2, -1/2$, correspondingly.:
\begin{equation}\label{HZ32}
\hat{H}_Z^{\text{eff},3/2}=\mu_B\left(
\begin{array}{cc}
 - \frac{3}{2}g_{||}^{3/2}B_z & -\frac{3}{2} g^{\perp}_{3/2} B_{+} \\
 -\frac{3}{2} g^{\perp}_{3/2} B_{-} &  \frac{3}{2}g^{||}_{3/2}B_z\\
\end{array}
\right),
\end{equation}
\begin{equation}\label{HZ12}
\hat{H}_Z^{\text{eff},1/2}=\mu_B\left(
\begin{array}{cc}
 - \frac{1}{2}g^{||}_{1/2}B_z & -\frac{1}{2}g^{\perp}_{1/2}B_{-} \\
- \frac{1}{2}g^{\perp}_{1/2}B_{+} &  \frac{1}{2}g^{||}_{1/2}B_z \\
\end{array}
\right).
\end{equation}
In that case each pair of states corresponding to the fixed $M$ can be described by pseudospin $1/2$.
In a spherically-symmetric system, $g^{\perp}_{3/2}=Q^{\rm eff}=0$ and $g^{||}_{3/2}=g^{||}_{1/2}=g^{\perp}_{1/2}/2=g_h$.

\section{Numerical calculation of hole energies,  wave functions and $g$ factors in spherical NCs with infinite box-like potential}
 \label{AB}

\setcounter{equation}{0}
\setcounter{figure}{0}
\setcounter{table}{0}
\renewcommand{\thefigure}{B\arabic{figure}}
\renewcommand{\theequation}{B\arabic{equation}}
\renewcommand{\thetable}{B\arabic{table}} 

We developed a numerical method for calculating hole energies, wave functions, and $g$ factors in spherical NCs with infinite box-like potentials, which is similar to our methods for spherical NCs with parabolic potential \cite{Semina2016} and cube-shaped NCs with infinite box-like potentials \cite{Semina2021}. It consists of  diagonalizing the hole Hamiltonian matrix, calculated
on  the basis of eigenfunctions of the Luttinger Hamiltonian in spherical approximation in spherical NCs with infinite box-like potentials for a given set of Luttinger parameters $\gamma_1$, $\gamma_2$ and $\gamma_3$ in zero magnetic field \cite{Gelmont1971} for full momentum $j$ and its projection on the $z$-axis \eqref{Gelmont_functions}.
For even solutions, the wave function for given values of $j$ and $M$ contains two contributions, with radial functions corresponding to $l$ equal to $j - 3/2$ and $j + 1/2$, respectively:  \cite{Ekimov1984} (we omit here the normalization constants) 
\begin{multline}
 R_{j,j-3/2}(r) =   \sqrt{\frac{2 j+3}{6 j-3}} j_{j-\frac{3}{2}}(k r)-\\-\frac{\sqrt{\frac{2 j+3}{6 j-3}} j_{j-\frac{3}{2}}(k) j_{j-\frac{3}{2}}\left(k r
   \sqrt{\beta }\right)}{j_{j-\frac{3}{2}}\left(k \sqrt{\beta }\right)},
\end{multline}
\begin{multline}
 R_{j,j+1/2}(r) =\\ =   \frac{(2 j+3) j_{F-\frac{3}{2}}(k) j_{j+\frac{1}{2}}\left(k r \sqrt{\beta }\right)}{(6 j-3) j_{j-\frac{3}{2}}\left(k \sqrt{\beta
   }\right)}+j_{j+\frac{1}{2}}(k r),
\end{multline}
where  $\beta=(\gamma_1-2\gamma)/(\gamma_1+2\gamma)$ is the light to heavy hole effective mass ratio and $j_{l}(x)$ are spherical Bessel functions. 
The equation determining the hole wave vector $k$ comes from the boundary condition of the wave function vanishing at the edge of the NC of radius $a$:
\begin{multline}
j_{j+\frac{1}{2}}(ka) j_{j-\frac{3}{2}}\left(\sqrt{\beta } ka\right) + \\+\frac{(2 j+3) j_{j-\frac{3}{2}}(ka)
   j_{j+\frac{1}{2}}\left(\sqrt{\beta } ka\right)}{6 j-3}=0
\end{multline}

For the odd states, the wave function for given values of $j$ and $M$ contains two contributions with $l$ being equal to  $j +3/2$ and $j - 1/2$ and one has 
\begin{multline}
 R_{j,j-1/2}(r) =   \sqrt{2 j-1} j_{j-\frac{1}{2}}(k r)-\\-\frac{\sqrt{2 j-1} j_{j-\frac{1}{2}}(k) j_{j-\frac{1}{2}}\left(k r \sqrt{\beta
   }\right)}{j_{j-\frac{1}{2}}\left(k \sqrt{\beta }\right)},
\end{multline}
\begin{multline}
 R_{j,j+3/2}(r) =  \frac{(1-2 j) j_{j+\frac{3}{2}}(k r)}{\sqrt{3} \sqrt{2 j+3}}-\\-\frac{\sqrt{3} \sqrt{2 j+3} j_{j-\frac{1}{2}}(k) j_{j+\frac{3}{2}}\left(k r
   \sqrt{\beta }\right)}{j_{j-\frac{1}{2}}\left(k \sqrt{\beta }\right)},
\end{multline}
And from the boundary conditions one has: 
\begin{multline}
(2j-1) j_{j+\frac{3}{2}}(ka) j_{j-\frac{1}{2}}\left(\sqrt{\beta } ka\right)+\\+3 (2 j+3)
   j_{j-\frac{1}{2}}(ka) j_{j+\frac{3}{2}}\left(\sqrt{\beta } ka\right)=0
\end{multline}
For the special case $j=1/2$ the only one radial function remains:
\begin{equation}
R_{1/2,1}(r) =j_1(k r)
\end{equation}
for even state and 
\begin{equation}
R_{1/2,2}(r) =j_2(k r)
\end{equation} 
for the odd state. And with simple boundary conditions 
\begin{equation}
  j_{1(2)}(k a)=0  
\end{equation}
for even (odd) states, correspondingly.

The introduced basis set is full, so one can use it for calculating hole states in a magnetic field.  In addition, it allows valence band warping to be taken into account. In real world calculations, we use a finite size basis set, while choosing it to be sufficiently large for needed accuracy.

\section{Trion emission lines in an applied magnetic field}
\label{AC}

\setcounter{equation}{0}
\setcounter{figure}{0}
\setcounter{table}{0}
\renewcommand{\thefigure}{C\arabic{figure}}
\renewcommand{\theequation}{C\arabic{equation}}
\renewcommand{\thetable}{C\arabic{table}} 

For a magnetic field oriented at an arbitrary angle $\theta$ with respect to the $z$-axis of a NC, energies of the negative trion spin states are the eigenenergies of the hole Hamiltonian \ref{Hzeff}. After  the trion recombination, there remains an electron with a spin projection on the magnetic field direction $\pm1/2$. In the case of the isotropic electron $g$ factor, electron energy does not depend on the angle $\theta$ and equals $\pm g_e\mu_BB/2$. Knowing energies of the excited and ground states, we can calculate energies for all possible trion emission lines. 

To calculate the relative oscillator strength of the trion emission lines, we calculate the eigenvectors of the excited and ground states of a NC using the set of basis spin states along the $z$-axis of the laboratory frame. The excited state of the system is described by eigenvectors of the Hamiltonian \ref{Hzeff}. The  dependence of the eigenvectors on the angle $\theta$ is described by the transformation matrix for spin $3/2$ with good accuracy. For the ground state of the system, we use the transformation matrix for spin $1/2$.
\begin{eqnarray}
    \ket{i}=\sum_{j=1}^4 T^{3/2}_{i,j}(\theta)\ket{j}, \\
    \ket{i'}=\sum_{j'=1}^2 T^{1/2}_{i',j'}(\theta)\ket{j'}
\end{eqnarray}
where $i,j=1,2,3,4$ correspond to hole states with angular momentum projection $M=3/2,1/2,-1/2,-3/2$ on the magnetic field direction and $z$-axis, respectively. $i',j'=1,2$ correspond to electron states with spin projection $S=1/2,-1/2$ on the magnetic field direction and $z$-axis, respectively. The matrices $T^{3/2}$ and $T^{1/2}$ are:

\begin{widetext}
\small{\begin{equation}\label{T32matrix}
T^{3/2}=\left(
\begin{array}{cccc}
 \cos^3(\theta/2)& \sqrt{3}\cos^2(\theta/2)\sin(\theta/2) & \sqrt{3}\cos(\theta/2)\sin^2(\theta/2)  &\sin^3(\theta/2)\\
 -\sqrt{3}\cos^2(\theta/2)\sin(\theta/2)& \cos(\theta/2)(2-3\cos^2(\theta/2)) & \sin(\theta/2)(2-3\sin^2(\theta/2)) & \sqrt{3}\cos(\theta/2)\sin^2(\theta/2)  \\
 \sqrt{3}\cos(\theta/2)\sin^2(\theta/2)  &  -\sin(\theta/2)(2-3\sin^2(\theta/2)) & \cos(\theta/2)(2-3\cos^2(\theta/2)) &
   \sqrt{3}\cos^2(\theta/2)\sin(\theta/2) \\
 -\sin^3(\theta/2) &\sqrt{3}\cos(\theta/2)\sin^2(\theta/2) & -\sqrt{3}\cos^2(\theta/2)\sin(\theta/2) &  \cos^3(\theta/2)\\
\end{array}
\right).
\end{equation}}
\normalsize \end{widetext}
\begin{equation}\label{T12}
T^{1/2}=\left(
\begin{array}{cc}
 \cos(\theta/2)& \sin(\theta/2) \\
- \sin(\theta/2) & \cos(\theta/2) \\
\end{array}
\right).
\end{equation}
For a photon emitted along the $z$-axis of the laboratory frame, we can calculate the relative oscillator strength of transitions between $\ket{i}$ and $\ket{i'}$ states. For this purpose, we use the following ratio between squared matrix elements of the operator $\bm{Ap}$, calculated on basis states along the $z$-axis of NC : 
\begin{eqnarray}
    |\langle j=\pm3/2| \bm{Ap} \ket{j'=\pm1/2}|^2\propto3, \nonumber \\ \nonumber
    |\bra{j=\pm1/2} \bm{Ap} \ket{j'=\pm1/2}|^2\propto2, \\ \nonumber
    |\bra{j=\pm1/2} \bm{Ap} \ket{j'=\mp 1/2}|^2\propto1, \\
    |\bra{j=\pm3/2} \bm{Ap} \ket{j'=\mp 1/2}|^2\propto0.
\end{eqnarray}
Here, $\bm A$ is the vector potential of electromagnetic field, ${\bm p}$ is the momentum operator.
Note that these selection rules are written for a negative trion with ground states composed of a resident electron. In the case of an electron-hole pair, the sign of the electron spin projection should be inverted.

The resulting intensities of the trion emission lines are calculated as:
\begin{eqnarray}
    I_{i,i'}=\exp\left(-\frac{(E-\Delta E_{i,i'})^2}{2\sigma^2}-\frac{E_i}{kT}\right)|\bra{i}Ap\ket{i'}|^2.
\end{eqnarray}
Here, the exponential prefactor takes into account broadening of trion lines with $\sigma=0.05$~meV and Boltzman population of initial trion states with energies $E_i$.

\end{document}